\documentclass[12pt]{article}

\usepackage{axodraw,graphicx,epsfig}
\usepackage[text={400pt,680pt},centering]{geometry}
\renewcommand{\theequation}{\thesection.\arabic{equation}}

\newcommand{\<}{\langle}
\renewcommand{\>}{\rangle}
\newcommand{\beq}{\begin{equation}}
\newcommand{\eeq}{\end{equation}}
\newcommand{\beqn}{\begin{eqnarray}}
\newcommand{\eeqn}{\end{eqnarray}}
\newcommand{\dd}{\bar{d}}

\def\simge{\mathrel{\rlap{\raise 0.511ex \hbox{$>$}}{\lower 0.511ex
\hbox{$\sim$}}}}
\def\simle{\mathrel{\rlap{\raise 0.511ex \hbox{$<$}}{\lower 0.511ex \hbox{$\sim$}}}}
\def\sla#1{\setbox0=\hbox{$#1$}\dimen0=\wd0
      \setbox1=\hbox{/} \dimen1=\wd1 \ifdim\dimen0>\dimen1
      \rlap{\hbox to \dimen0{\hfil/\hfil}} #1                        \else
      \rlap{\hbox to \dimen1{\hfil$#1$\hfil}}
      /   \fi}
\newcommand{\nn}{\nonumber}
\newcommand{\ov}{\overline}
\newcommand{\ra}{\rightarrow}
\newcommand{\mc}{\mathcal}
\newcommand{\FA}{\textsf{FeynArts}}
\newcommand{\Math}{\textsf{Mathematica}}

\newcommand{\as}{\frac{\alpha_s}{4\pi}}
\newcommand{\al}{\alpha}
\newcommand{\be}{\beta}
\newcommand{\ga}{\gamma}
\newcommand{\de}{\delta}
\newcommand{\Ga}{\Gamma}

\newcommand{\vep}{\varepsilon}
\newcommand{\la}{\lambda}

\newcommand{\tilga}{\tilde{\gamma}}

\newcommand{\tilg}{\tilde{g}}
\newcommand{\ie}{\emph{i.e.} }
\newcommand{\eg}{\emph{e.g.} }
\newcommand{\msbar}{\overline{\mbox{\scriptsize MS}}}
\newcommand{\MSbar}{\overline{\mbox{MS}}}

\begin{document}
\thispagestyle{empty}
\begin{flushright}
\vskip 0.5cm
\begin{tabular}{l}
{\tt RM3-TH/06-10}\\
\end{tabular}
\end{flushright}

\vspace{1.0cm}

\begin{center}
\LARGE {\bf \boldmath Next-to-leading order strong\\
\vspace{0.2cm} interaction corrections to the $\Delta F=2$\\
\vspace{0.2cm} effective Hamiltonian in the MSSM}

\vspace{1.5cm}

\noindent {\sc \large M.~Ciuchini$^{\,a}$, E.~Franco$^{\,b}$,
D.~Guadagnoli$^{\,c}$,\\
V.~Lubicz$^{\,a}$, V.~Porretti$^{\,a}$, L.~Silvestrini$^{\,b}$}

\vspace{1.0cm}

\normalsize
\noindent {\sl $^a$ Dipartimento di Fisica, Universit\`{a} di Roma Tre and
INFN, \\ Via della Vasca Navale 84, I-00146 Rome, Italy \\
$^b$ Dipartimento di Fisica, Universit\`{a} di Roma ``La Sapienza'' and
INFN, \\ P.le A. Moro 2, I-00185 Rome, Italy \\
$^c$ Technische Universit{\"a}t M{\"u}nchen, Physik Department, \\
D-85748 Garching, Germany}

\end{center}

\vspace{1.5cm}
\begin{abstract}
We compute the next-to-leading order strong interaction corrections to
gluino-mediated $\Delta F=2$ box diagrams in the Minimal Supersym\-me\-tric
Standard Model. These corrections are given by two loop diagrams which we
have calculated in three different regularization schemes in the mass
insertion approximation. We obtain the next-to-leading order Wilson
coefficients of the $\Delta F=2$ effective Hamiltonian relevant for neutral
meson mixings. We find that the matching scale uncertainty is largely
reduced at the next-to-leading order, typically from about 10-15\% to few
percent.
\end{abstract}

\newpage
\noindent \line(1,0){400}
\vspace{-0.3cm}
\tableofcontents
\noindent \line(1,0){400}

\section{Introduction\label{sec:intro}}
\setcounter{equation}{0} 

If ongoing and planned experiments discover direct or indirect departures
from the Standard Model (SM), the next step will be to understand what kind
of new physics is involved. Detailed predictions for flavour changing
neutral currents processes in supersymmetry play a crucial role in this
program. In particular, flavour mixing induced by sfermion mass matrices is
a pure supersymmetric effect with no analogue in the SM and constitutes in
general the bulk of SUSY contributions to neutral meson mixings. These
processes provide in turn one of the most sensitive guideline for
reconstructing the structure of SUSY soft breaking terms. Ultimately, this
information will allow us to discriminate among the many possible mechanisms
for SUSY breaking that have been proposed in the literature.

In this paper we present the next-to-leading order (NLO) strong interaction
corrections to gluino-mediated $\Delta F=2$ box diagrams in the Minimal
Supersymmetric Standard Model (MSSM). We obtain the NLO Wilson coefficients
of the effective Hamiltonian relevant for neutral meson mixings. We adopt
the mass insertion approximation~\cite{mia} which is phenomenologically
motivated and permits a compact presentation of the results for the Wilson
coefficients.

The complete expressions of the Wilson coefficients at the NLO are collected
in appendix~\ref{app:coe}, where the results are presented in the
$\MSbar$-DRED renormalization scheme. In eqs.~(\ref{DZnd})-(\ref{NDRtoDRED})
and (\ref{Dr-dredri})-(\ref{DREDtoRI}) we provide the formulae required to
translate the Wilson coefficients to the $\MSbar$-NDR and RI-MOM schemes.
The relations between the strong coupling constant and the squark and gluino
masses in the $\MSbar$-DRED and NDR schemes are given in eq.~(\ref{asdred}).

At the LO, strong interaction contributions to $\Delta F=2$ processes are
described in SUSY by the gluino mediated box diagrams represented in
fig.~\ref{fig:diagsLO}. These diagrams have been computed in
refs.~\cite{Gerard:1984}-\cite{GGMS}. NLO corrections to $\Delta F=2$
processes are available for the chargino contributions in the
MSSM~\cite{Krauss:1998mt} and for the Two Higgs Doublet
Model~\cite{Urban:1997gw}. Both chargino and gluino contributions have been
then computed in~\cite{Feng:2000kg} in the MSSM with minimal flavour
violation. The anomalous dimension matrix for the complete set of
four-fermion operators entering the effective $\Delta F=2$ Hamiltonian has
been evaluated at the NLO in QCD in refs.~\cite{NLOADM,NLOADMcheck}.

This paper completes the NLO determination of the effective Hamiltonian by
computing the initial conditions for the Wilson coefficients at the
high-energy supersymmetric scale. Besides the general argument that initial
conditions are needed to obtain scheme-independent results and to achieve
NLO scale invariance, our calculation is strongly motivated by two
additional considerations. First, the LO coefficients generated by gluino
exchange are proportional to $\alpha_s^2$. Without the NLO computation of
matching conditions, it is not possible to specify the scale and scheme for
the strong coupling, resulting in an uncertainty of the LO result much
larger than in ordinary weak-interaction processes. Second, the new $\Delta
F=2$ operators generated by gluino exchange have surprisingly large
anomalous dimensions, so that there is a large scale dependence that can
only be removed by adding the NLO corrections to the matching (see
eq.~(\ref{chain-mu-dep})). We consider two different regularization schemes
for ultraviolet (UV) divergences, namely the naive dimensional
regularization (NDR) and the dimensional reduction (DRED). Infrared (IR)
divergences are treated both with a gluon mass (in the NDR and DRED schemes)
and with dimensional regularization (in DRED). The main achievement of the
NLO determination is a strong reduction of the high-energy scale dependence
of the Wilson coefficients compared to the LO, typically from about 10-15\%
to few percent. Applications of our calculation are studies of $B_{d,s}-\bar
B_{d,s}$, $D-\bar D$ and $K-\bar K$ mixings. Preliminary results for the
$B_{d}-\bar B_{d}$ mixing case have been given in ref.~\cite{epsdiego} and a
complete phenomenological analysis will be presented in a forthcoming paper.

The plan of the paper is the following. In section~\ref{sec:EffHam} we
introduce the effective Hamiltonian approach and the basic formulae used in
the matching procedure at the NLO. In section~\ref{sec:NLOfull} we discuss
the calculation in the full theory (the MSSM) both at the LO and at the NLO.
The latter represents the main result of the paper. We give details of the
calculation and address in particular the issues related to the role of
evanescent operators in the matching. In section~\ref{sec:NLOeff} we present
the calculation in the effective theory. The results for the Wilson
coefficients are discussed in section~\ref{sec:NLOmatch} together with the
consistency checks between results obtained in the different UV and IR
regularization schemes and the scaling under the renormalization group
equation. Finally, in section~\ref{sec:conclusions}, we draw our
conclusions. The complete expressions of the Wilson coefficients, both at
the LO and at the NLO, are collected in appendix~\ref{app:coe} 

\section{Effective Hamiltonian for \boldmath $\Delta F=2$
processes\label{sec:EffHam}} 
\setcounter{equation}{0} 

The effective Hamiltonian for $\Delta F=2$ processes in the presence of
new physics can be written in terms of eight independent four-fermion
operators,
\beq
{\cal H}_{\rm eff}^{\Delta F=2}=\sum_{i=1}^{5} C_i\, {\cal O}_i +
\sum_{i=1}^{3} \tilde{C}_i\, \tilde{{\cal O}}_i \, , 
\label{heff} 
\eeq
where $C_i$ are the Wilson coefficients and we adopt the following
basis for the local operators ${\cal O}_i$ 
\beqn
 {\cal O}_1 & = & \dd^{i} \gamma_{\mu\,L}\, b^{i}\, \dd^{j} \gamma^\mu_{L}\,
b^{j}\; ,\nn \\
 {\cal O}_2 & = & \dd^{i} P_{L}\, b^{i}\, \dd^{j} P_{L}\, b^{j}\; ,\nn \\
 {\cal O}_3 & = & \dd^{i} P_{L}\, b^{j}\, \dd^{j} P_{L}\, b^{i}\; ,\nn\\
 {\cal O}_4 & = & \dd^{i} P_{L}\, b^{i}\, \dd^{j} P_{R}\, b^{j}\; ,\nn \\
 {\cal O}_5 & = & \dd^{i} P_{L}\, b^{j}\, \dd^{j} P_{R}\, b^{i}\; .
 \label{basis}
\eeqn
The operators $\tilde{{\cal O}}_{1,2,3}$ are obtained from ${\cal
O}_{1,2,3}$ by the exchange $L\leftrightarrow R$. The left- and right-handed
projectors are defined as $P_{R,L}= (1\pm\gamma_5)/2$ and
$\gamma^\mu_{R,L}=\gamma^\mu P_{R,L}$; $i,j$ are colour indices. In
eq.~(\ref{basis}) and in the following we specialized for definiteness on
the effective Hamiltonian which describes $\bar B_d -B_d$ mixing. In the
case of $B_s$, $D$ and $K$ mixings, the replacements $\{d,b\}\rightarrow
\{s,b\}$, $\{d,b\}\rightarrow \{u,c\}$ and $\{d,b\}\rightarrow \{d,s\}$
should be considered respectively.

The evaluation of the coefficients of an effective Hamiltonian involves the
following two steps:
\begin{enumerate}
\item calculating the amplitude in both the full and the effective theory
and determining the Wilson coefficients by matching the two amplitudes at
the high energy scale;
\item evolving the Wilson coefficients from the high- to the low-energy
scale where the matrix elements of the local operators can be computed with
non-perturbative methods, primarily lattice QCD calculations.
\end{enumerate}
Step 1 depends on the theory under consideration. The new result of this
paper is the computation of the full theory amplitude in the MSSM up to the
NLO in the strong interactions. As far as step 2 is concerned, the NLO
anomalous dimension of the effective Hamiltonian in eq.~(\ref{heff}) has
been calculated in ref.~\cite{NLOADM} and the result confirmed
in~\cite{NLOADMcheck}.

We now recall the general formulae necessary to perform the matching
between the full and the effective theories at the NLO.

The renormalized amplitude in the full theory can be written in the form
\beqn 
\mc{A}_{\rm full} = \sum_i \left( F^{(0)}_i + \as F^{(1)}_i
\right) \< \mc{O}_i \>^{(0)}~, 
\label{Afull-form} 
\eeqn
where $\< \mc{O}_i \>^{(0)}$ are the tree level matrix elements of the
operators $\mc{O}_i$ and $F^{(0)}$ and $F^{(1)}$ represent the LO and NLO
contributions respectively. Note that, in the case of the $\Delta F=2$ SUSY
transitions considered in this paper, both $F^{(0)}$ and $F^{(1)}$ contain
an additional factor $\al_s^2$ not factorized out in eq.~(\ref{Afull-form}).
It is also worth recalling that, in order to properly normalize the physical
amplitude, the external quark fields considered to compute the amplitudes
should be renormalized with their on-shell renormalization constant, defined
as the pole residue of the quark propagator. In the calculation performed in
this paper the external fields, as well as the strong coupling constant,
renormalize differently in the full (MSSM) and in the effective theory, and
this gives a finite contribution to the matching. In particular, one loop
corrections to the quark propagator in the full theory include a
squark-gluino loop as well as a quark-gluon loop, whereas only the latter
appears in the low-energy effective theory.

It is convenient to express also the NLO renormalized amplitude in the
effective theory in terms of tree-level matrix elements of local operators,
\beqn \mc{A}_{\rm eff} = \sum_i C_i \< \mc{O}_i \> =
\sum_{i,j} C_i \left( 1 + \as r\right)_{ij} \< \mc{O}_j \>^{(0)}~.
\label{Aeff-form} 
\eeqn 
By equating the full theory amplitude in eq.~(\ref{Afull-form}) with the
effective one given in eq.~(\ref{Aeff-form}) one obtains the expression for
the Wilson coefficients at the NLO,
\beqn 
C_j = F^{(0)}_j + \as F^{(1)}_j - \as \sum_k F^{(0)}_k r_{kj}~.
\label{match-passage} 
\eeqn

The functions $F^{(i)}$ and $r$ depend in general on the external states. In
our calculation we have chosen massless external quarks with zero momenta.
Though this choice considerably simplifies the calculation of the two-loop
diagrams in the full theory, it also introduces IR divergences in both the
full and effective theories, in particular in $F^{(1)}$ and $r$. These
divergences cancel in the Wilson coefficients. Particular care, however,
must be taken when regularizing IR divergences in dimensional
regularization. In this case, the matrix $r$ contains $1/\epsilon$ poles
that give finite contributions to eq.~(\ref{match-passage}) once combined
with both $\mc{O}(\epsilon)$-terms entering $F^{(0)}$ and contributions to
$F^{(0)}$ of evanescent operators. In particular, the summation index $k$ in
eq.~(\ref{match-passage}) must run in this case over both the physical and
the evanescent operators, whose specific definition will be given in the
next section. The evanescent operators which are needed instead to define
the renormalization scheme of four-fermion operators within dimensional
regularization are discussed in sec.~\ref{sec:NLOeff}.

\section{Calculation in the full theory\label{sec:NLOfull}}
\setcounter{equation}{0} 

We now describe the NLO calculation of the Wilson coefficients for $\Delta
F=2$ transitions mediated by strong interactions in the MSSM. We will
discuss in turn the computation of all the elements entering the r.h.s. of
eq.~(\ref{match-passage}): the determination of the LO and NLO amplitudes in
the full theory, $F^{(0)}$ and $F^{(1)}$, is discussed in this section; the
calculation of the amplitude in the effective theory, expressed by matrix
$r$, will be discussed in section \ref{sec:NLOeff}.

As mentioned before, having chosen external quarks with zero masses and
momenta, the bare amplitudes in both the full and effective theories present
UV as well as IR divergences. To regularize both of them we have adopted
three regularization setups:
\begin{itemize}
\item DRED, with a gluon mass $\la$ as IR regulator (DRED-$\la$);
\item DRED, to regularize both UV and IR divergences (DRED-$d$);
\item NDR, with a gluon mass $\la$ (NDR-$\la$).
\end{itemize}
The calculation is performed in the mass insertion approximation \cite{mia}
which is phenomenologically motivated and allows a more compact presentation
of the final results. In order to fix the notation, we recall here the basic
formula of the mass insertion approximation which provides the expansion of
the squark mass matrix in the flavour basis around its mean diagonal value,
\beqn 
(Z^\dagger)_{ik}(M_D^2)_{k} (Z)_{kj} =
(M^2)_{ij} = ~M_s^2 \left(1 + \frac{\Delta}{M_s^2} \right)_{ij} = M_s^2
\left(1 + \delta\right)_{ij} ~. 
\label{app:miaeq}
\eeqn
The matrices $Z$ and $M_D$ are the squark mixing and mass matrix
respectively in the mass eigenstate basis; $M$ is the squark mass matrix in
the super-CKM basis $(\,\tilde q^1_L\, \tilde q^2_L \,\tilde q^3_L\, \tilde
q^1_R\, \tilde q^2_R \,\tilde q^3_R\,)$; $M_s$ is a mean squark mass, as
defined for example in \cite{GGMS}; $\Delta_{ij}$ ($\delta_{ij}$) are the
dimensionful (dimensionless) mass insertions between squarks of flavour $i$
and $j$. We treat $M_S$ as the usual mass parameter in the Lagrangian and
the $\delta$'s as interaction terms. We then expand the $\Delta F=2$
amplitude up to the second order in the $\delta$'s, which provides the first
non-vanishing contribution in the mass insertion approximation.

\subsection{LO calculation up to $\mathbf{\mc{O}(\epsilon)}$\label{LO}}

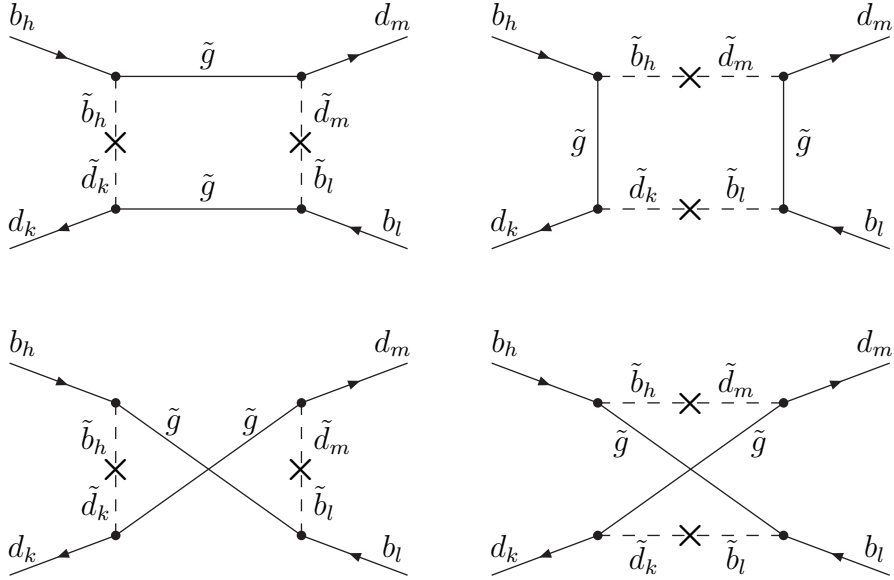
\begin{figure}
\vspace{-0.3cm}
\begin{center}
\SetScale{1}
\begin{picture}(150,80)(0,0)
\Vertex(40,15){1.8} \Vertex(40,65){1.8} \Vertex(110,15){1.8}
\Vertex(110,65){1.8}
\Text(5,10)[c]{$d_k$}
\Text(5,88)[c]{$b_h$}
\Text(145,10)[c]{$b_l$}
\Text(145,88)[c]{$d_m$}
\ArrowLine(40,15)(0,0) \ArrowLine(0,80)(40,65)
\ArrowLine(150,0)(110,15) \ArrowLine(110,65)(150,80)
\Text(75,73)[c]{$\tilde{g}$}
\Text(75,23)[c]{$\tilde{g}$}
\Line(40,65)(110,65) \Line(110,15)(40,15)
\Text(110,40)[c]{\large \boldmath $\times$}\Text(40,40)[c]{\large \boldmath
$\times$}
\Text(27,52.5)[l]{$\tilde{b}_h$}
\Text(27,27.5)[l]{$\tilde{d}_k$}
\Text(115,52.5)[l]{$\tilde{d}_m$}
\Text(115,27.5)[l]{$\tilde{b}_l$}
\DashLine(110,15)(110,65){5} \DashLine(40,65)(40,15){5}
\end{picture}
\hspace{1cm}
\begin{picture}(150,80)(0,0)
\Vertex(40,15){1.8} \Vertex(40,65){1.8} \Vertex(110,15){1.8}
\Vertex(110,65){1.8}
\Text(5,10)[c]{$d_k$}
\Text(5,88)[c]{$b_h$}
\Text(145,10)[c]{$b_l$}
\Text(145,88)[c]{$d_m$}
\ArrowLine(40,15)(0,0) \ArrowLine(0,80)(40,65)
\ArrowLine(150,0)(110,15) \ArrowLine(110,65)(150,80)
\DashLine(40,65)(110,65){5} \DashLine(110,15)(40,15){5}
\Text(75,15)[c]{\large \boldmath $\times$}\Text(75,65)[c]{\large \boldmath
$\times$}
\Text(57.5,74)[c]{$\tilde{b}_h$}
\Text(92.5,74)[c]{$\tilde{d}_m$}
\Text(57.5,24)[c]{$\tilde{d}_k$}
\Text(92.5,24)[c]{$\tilde{b}_l$}
\Line(110,15)(110,65) \Line(40,15)(40,65)
\Text(30,40)[l]{$\tilde{g}$}
\Text(115,40)[l]{$\tilde{g}$}
\end{picture}

\vspace{1.5cm}

\begin{picture}(150,80)(0,0)
\Vertex(40,15){1.8} \Vertex(40,65){1.8} \Vertex(110,15){1.8}
\Vertex(110,65){1.8}
\Text(5,10)[c]{$d_k$}
\Text(5,88)[c]{$b_h$}
\Text(145,10)[c]{$b_l$}
\Text(145,88)[c]{$d_m$}
\ArrowLine(40,15)(0,0) \ArrowLine(0,80)(40,65)
\ArrowLine(150,0)(110,15) \ArrowLine(110,65)(150,80)
\Text(110,40)[c]{\large \boldmath $\times$}\Text(40,40)[c]{\large \boldmath
$\times$}
\Text(27,52.5)[l]{$\tilde{b}_h$}
\Text(27,27.5)[l]{$\tilde{d}_k$}
\Text(115,52.5)[l]{$\tilde{d}_m$}
\Text(115,27.5)[l]{$\tilde{b}_l$}
\DashLine(110,15)(110,65){5} \DashLine(40,65)(40,15){5}
\Text(62,58)[c]{$\tilde{g}$}
\Text(91,58)[c]{$\tilde{g}$}
\Line(40,65)(110,15) \Line(110,65)(40,15)
\end{picture}
\hspace{1cm}
\begin{picture}(150,80)(0,0)
\Vertex(40,15){1.8} \Vertex(40,65){1.8} \Vertex(110,15){1.8}
\Vertex(110,65){1.8}
\Text(5,10)[c]{$d_k$}
\Text(5,88)[c]{$b_h$}
\Text(145,10)[c]{$b_l$}
\Text(145,88)[c]{$d_m$}
\ArrowLine(40,15)(0,0) \ArrowLine(0,80)(40,65)
\ArrowLine(150,0)(110,15) \ArrowLine(110,65)(150,80)
\DashLine(40,65)(110,65){5} \DashLine(110,15)(40,15){5}
\Text(75,15)[c]{\large \boldmath $\times$}\Text(75,65)[c]{\large \boldmath
$\times$}
\Text(57.5,74)[c]{$\tilde{b}_h$}
\Text(92.5,74)[c]{$\tilde{d}_m$}
\Text(57.5,7)[c]{$\tilde{d}_k$}
\Text(92.5,7)[c]{$\tilde{b}_l$}
\Text(49,50)[c]{$\tilde{g}$}
\Text(101,50)[c]{$\tilde{g}$}
\Line(110,15)(40,65) \Line(40,15)(110,65)
\end{picture}
\end{center}
\caption{\small\sl Feynman diagrams describing the gluino contribution to
the $B_d-\bar B_d$ transition in the MSSM. A cross indicates a mass
insertion and the indices $h,k,l,m$ label the squark chiralities. The
diagrams denoted as A-type and B-type in the text are those represented in
the first and second row respectively.}
\label{fig:diagsLO}
\end{figure}

The amplitude of $\Delta F=2$ transitions via strong interactions at the LO
in the MSSM receives contribution from the four box diagrams represented in
fig. \ref{fig:diagsLO} for the $B_d-\bar B_d$ mixing case.

We denote the diagrams in the first and second row of fig. \ref{fig:diagsLO}
as A-type and B-type diagrams respectively and we will extend this notation
to the analogous topologies entering at the NLO as well (see
fig.~\ref{fig:diagsNLO-gluon}). B-type diagrams entail the typical ambiguity
in defining the fermion flow present when dealing with Majorana fermions.
For a discussion on this point and for the Feynman rules of the MSSM we
refer the reader to the refs.~\cite{Haber:1984rc}-\cite{DEHK}.

According to eq.~(\ref{match-passage}), the Wilson coefficients at the LO
are given directly by the amplitudes $F^{(0)}_j$. As discussed in the
previous section, however, in the presence of dimensionally regularized IR
divergences, the NLO calculation of the Wilson coefficients also requires
the evaluation of the LO coefficients of the physical operators up to
${\mc{O}(\epsilon)}$, as well as the evaluation at the LO of the
coefficients of the evanescent operators. This is due to the presence of the
last term in eq.~(\ref{match-passage}). In the DRED regularization scheme,
one finds the appearance of both a $d$-dimensional metric tensor
$g_{\mu\nu}$ generated by loop integration (the momenta are $d$-dimensional)
and of a four-dimensional tensor, $\tilg_{\mu\nu}$, coming from the algebra
of four-dimensional gamma matrices. Evanescent operators are generated in
this scheme by the contraction of Dirac strings with the tensor $\Delta
g_{\mu\nu}$, which can be defined by the following splitting of the metric
tensor~\cite{BW}:
\beqn 
g_{\mu \nu} = \frac{d}{4} \tilg_{\mu \nu} + \left( g_{\mu \nu} - \frac{d}{4}
\tilg_{\mu \nu} \right) \equiv\frac{d}{4} \tilg_{\mu \nu} + \Delta
g_{\mu\nu}~,
\label{app:gsplit} 
\eeqn 
where $d=4-2\vep$ and the relations
\beqn g_{\mu \nu} ~ \tilg^\nu_\rho = g_{\mu \rho} \quad , \quad \Delta
g_{\mu\nu} ~ \tilg^{\mu\nu}=0
\label{dred} 
\eeqn
define the contraction rules in the DRED scheme. The term $\Delta
g_{\mu\nu}$ is of $\mc{O}(\epsilon)$ and provides our definition of the
evanescent operators. In the calculation of the LO diagrams we find the
appearance of the following evanescent operators:
\beqn E^{\rm DRED}_1 &=& \Delta g_{\mu\nu}\;
 \bar{d}^i {\tilga}^{\mu}_L b^i \, \bar{d}^j {\tilga}^{\nu}_L b^j ~, \nn \\
E^{\rm DRED}_2 &=& \Delta g_{\mu\nu}\;
 \bar{d}^i {\tilga}^{\mu}_L b^i\, \bar{d}^j {\tilga}^{\nu}_R b^j ~, \nn \\
E^{\rm DRED}_3 &=& \Delta g_{\mu\nu}\;
 \bar{d}^i {\tilga}^{\mu}_L b^j \, \bar{d}^j {\tilga}^{\nu}_R b^i ~,
\label{basis-ev}
\eeqn
plus $\tilde{E}^{\rm DRED}_1$, obtained from $E_1^{\rm DRED}$ via the
exchange $L \leftrightarrow R$.

We have performed the LO calculation by using the three regularization
schemes discussed at the beginning of this section. The scheme independent
results for the Wilson coefficients of the physical operators at the LO, in
four dimensions, are in agreement with those obtained in ref.~\cite{GGMS}
and are presented for completeness in appendix~\ref{app:coe}.

\subsection{NLO calculation \label{NLO}}
The Feynman diagrams entering the calculation of the amplitude at the NLO
are shown in figs.~\ref{fig:diagsNLO-gluon}-\ref{fig:diagsNLO-X}.
\begin{figure}[p]
\begin{tabular*}{14.2cm}{cc@{\hspace{1.5cm}}cc@{\hspace{1.5cm}}c}
{\bf A-type} & {\bf graph} & {\bf B-type} & {\bf graph} & {\bf \#}\\
[0.1cm]
\hline \hline
\\
\raisebox{24pt}{$A_{12}$} & \SetScale{0.2}
\begin{picture}(90,48)(0,0) 
\SetWidth{2.5}
\SetColor{Black}
\Vertex(120,45){6}
\Vertex(120,195){6}
\Vertex(330,45){6}
\Vertex(330,195){6}
\ArrowLine(120,45)(0,0)
\ArrowLine(450,0)(330,45)
\Line(120,195)(330,195)
\Line(330,45)(120,45)
\DashArrowLine(330,45)(330,195){10}
\DashArrowLine(120,195)(120,45){10}
\ArrowLine(0,240)(50,220)
\ArrowLine(50,220)(120,195)
\Vertex(50,220){6}
\Vertex(395,220){6}
\ArrowLine(330,195)(395,220)
\ArrowLine(395,220)(450,240)
\GlueArc(222.5,-1262.5)(1492.5,83.36,96.64){4}{23.14}
\end{picture} &
\raisebox{24pt}{$B_{12}$} & \SetScale{0.2}
\begin{picture}(90,48) (0,0)
\SetWidth{2.5}
\SetColor{Black}
\Vertex(120,45){6}
\Vertex(120,195){6}
\Vertex(330,45){6}
\Vertex(330,195){6}
\ArrowLine(120,45)(0,0)
\ArrowLine(450,0)(330,45)
\Line(120,195)(330,45)
\Line(330,195)(120,45)
\DashArrowLine(330,45)(330,195){10}
\DashArrowLine(120,195)(120,45){10}
\Vertex(50,220){6}
\Vertex(400,220){6}
\ArrowLine(50,220)(120,195)
\ArrowLine(0,240)(50,220)
\ArrowLine(330,195)(400,220)
\ArrowLine(400,220)(450,240)
\GlueArc(225,-2840)(3065,86.73,93.27){4}{24}
\end{picture}
& \raisebox{24pt}{$4$}
\\
[0.11cm] \raisebox{24pt}{$A_{13}$} & \SetScale{0.2}
\begin{picture}(90,48)(0,0)
\SetWidth{2.5}
\SetColor{Black}
\Vertex(120,45){6}
\Vertex(120,195){6}
\Vertex(330,45){6}
\Vertex(330,195){6}
\ArrowLine(120,45)(0,0)
\ArrowLine(330,195)(450,240)
\Line(120,195)(330,195)
\Line(330,45)(120,45)
\DashArrowLine(330,45)(330,195){10}
\DashArrowLine(120,195)(120,45){10}
\Vertex(50,220){6}
\Vertex(400,20){6}
\ArrowLine(50,220)(120,195)
\ArrowLine(0,240)(50,220)
\ArrowLine(400,20)(330,45)
\ArrowLine(450,0)(400,20)
\GlueArc(115.48,-71.67)(298.93,17.86,102.65){4}{30}
\end{picture} &
\raisebox{24pt}{$B_{13}$} & \SetScale{0.2}
\begin{picture}(90,48) (0,0)
\SetWidth{2.5}
\SetColor{Black}
\Vertex(120,45){6}
\Vertex(120,195){6}
\Vertex(330,45){6}
\Vertex(330,195){6}
\ArrowLine(120,45)(0,0)
\Line(120,195)(330,45)
\Line(330,195)(120,45)
\DashArrowLine(330,45)(330,195){10}
\DashArrowLine(120,195)(120,45){10}
\Vertex(50,220){6}
\ArrowLine(50,220)(120,195)
\ArrowLine(0,240)(50,220)
\ArrowLine(330,195)(450,240)
\Vertex(400,20){6}
\ArrowLine(400,20)(330,45)
\ArrowLine(450,0)(400,20)
\GlueArc(111.83,-78.05)(304.39,18.79,101.72){4}{30}
\end{picture} &
\raisebox{24pt}{$4$}
\\
[0.11cm] \raisebox{24pt}{$A_{14}$} & \SetScale{0.2}
\begin{picture}(90,48) (0,0)
\SetWidth{2.5}
\SetColor{Black}
\Vertex(120,45){6}
\Vertex(120,195){6}
\Vertex(330,45){6}
\Vertex(330,195){6}
\ArrowLine(330,195)(450,240)
\Line(120,195)(330,195)
\Line(330,45)(120,45)
\DashArrowLine(330,45)(330,195){10}
\DashArrowLine(120,195)(120,45){10}
\Vertex(50,220){6}
\ArrowLine(50,220)(120,195)
\ArrowLine(0,240)(50,220)
\ArrowLine(450,0)(330,45)
\Vertex(50,20){6}
\ArrowLine(120,45)(50,20)
\ArrowLine(50,20)(0,0)
\GlueArc(290,120)(260,157.38,202.62){4}{12.86}
\end{picture} &
\raisebox{24pt}{$B_{14}$} & \SetScale{0.2}
\begin{picture}(90,48) (0,0)
\SetWidth{2.5}
\SetColor{Black}
\Vertex(120,45){6}
\Vertex(120,195){6}
\Vertex(330,45){6}
\Vertex(330,195){6}
\Line(120,195)(330,45)
\Line(330,195)(120,45)
\DashArrowLine(330,45)(330,195){10}
\DashArrowLine(120,195)(120,45){10}
\Vertex(50,220){6}
\ArrowLine(50,220)(120,195)
\ArrowLine(0,240)(50,220)
\ArrowLine(330,195)(450,240)
\ArrowLine(450,0)(330,45)
\Vertex(50,20){6}
\ArrowLine(120,45)(50,20)
\ArrowLine(50,20)(0,0)
\GlueArc(1047.5,120)(1002.5,174.28,185.72){4}{12.86}
\end{picture} &
\raisebox{24pt}{$4$}
\\
[0.11cm] \raisebox{24pt}{$A_{15}$} & \SetScale{0.2}
\begin{picture}(90,48)(0,2)
\SetWidth{2.5}
\SetColor{Black}
\Vertex(120,56){6}
\Vertex(120,206){6}
\Vertex(330,56){6}
\Vertex(330,206){6}
\ArrowLine(330,206)(450,251)
\Line(120,206)(330,206)
\Line(330,56)(120,56)
\DashArrowLine(330,56)(330,206){10}
\DashArrowLine(120,206)(120,56){10}
\Vertex(50,231){6}
\ArrowLine(50,231)(120,206)
\ArrowLine(0,251)(50,231)
\ArrowLine(450,11)(330,56)
\ArrowLine(120,56)(0,11)
\Vertex(200,206){6}
\GlueArc(113.25,148)(104.35,33.77,127.31){4}{11.14}
\end{picture} &
\raisebox{24pt}{$B_{15}$} & \SetScale{0.2}
\begin{picture}(90,48) (0,0)
\SetWidth{2.5}
\SetColor{Black}
\Vertex(120,45){6}
\Vertex(120,195){6}
\Vertex(330,45){6}
\Vertex(330,195){6}
\Line(120,195)(330,45)
\Line(330,195)(120,45)
\DashArrowLine(330,45)(330,195){10}
\DashArrowLine(120,195)(120,45){10}
\Vertex(50,220){6}
\ArrowLine(50,220)(120,195)
\ArrowLine(0,240)(50,220)
\ArrowLine(330,195)(450,240)
\ArrowLine(450,0)(330,45)
\ArrowLine(120,45)(0,0)
\Vertex(175,155){6}
\GlueArc(135,230.77)(85.68,-172.78,-62.17){4}{10.29}
\end{picture} &
\raisebox{24pt}{$8$}
\\
[0.11cm] \raisebox{24pt}{$A_{16}$} & \SetScale{0.2}
\begin{picture}(90,48) (0,0)
\SetWidth{2.5}
\SetColor{Black}
\Vertex(120,45){6}
\Vertex(120,195){6}
\Vertex(330,45){6}
\Vertex(330,195){6}
\ArrowLine(330,195)(450,240)
\Line(120,195)(330,195)
\Line(330,45)(120,45)
\DashArrowLine(330,45)(330,195){10}
\DashArrowLine(120,195)(120,45){10}
\Vertex(50,220){6}
\ArrowLine(50,220)(120,195)
\ArrowLine(0,240)(50,220)
\ArrowLine(450,0)(330,45)
\ArrowLine(120,45)(0,0)
\Vertex(195,45){6}
\GlueArc(280.99,263.82)(235.11,-169.26,-111.45){4}{15.43}
\end{picture} &
\raisebox{24pt}{$B_{16}$} & \SetScale{0.2}
\begin{picture}(90,48) (0,0)
\SetWidth{2.5}
\SetColor{Black}
\Vertex(120,45){6}
\Vertex(120,195){6}
\Vertex(330,45){6}
\Vertex(330,195){6}
\Line(120,195)(330,45)
\Line(330,195)(120,45)
\DashArrowLine(330,45)(330,195){10}
\DashArrowLine(120,195)(120,45){10}
\Vertex(50,220){6}
\ArrowLine(50,220)(120,195)
\ArrowLine(0,240)(50,220)
\ArrowLine(330,195)(450,240)
\ArrowLine(450,0)(330,45)
\ArrowLine(120,45)(0,0)
\Vertex(280,160){6}
\GlueArc(105.11,-39.57)(265.35,48.77,101.99){4}{16.29}
\end{picture} &
\raisebox{24pt}{$8$}
\\
[0.11cm] \raisebox{24pt}{$A_{17}$} & \SetScale{0.2}
\begin{picture}(90,48) (0,0)
\SetWidth{2.5}
\SetColor{Black}
\Vertex(120,45){6}
\Vertex(120,195){6}
\Vertex(330,45){6}
\Vertex(330,195){6}
\ArrowLine(330,195)(450,240)
\Line(120,195)(330,195)
\Line(330,45)(120,45)
\DashArrowLine(330,45)(330,195){10}
\Vertex(50,220){6}
\ArrowLine(50,220)(120,195)
\ArrowLine(0,240)(50,220)
\ArrowLine(450,0)(330,45)
\ArrowLine(120,45)(0,0)
\Vertex(120,120){6}
\DashArrowLine(120,195)(120,120){10}
\DashArrowLine(120,120)(120,45){10}
\GlueArc(135.82,205.57)(87.02,170.46,259.53){4}{8.57}
\end{picture} &
\raisebox{24pt}{$B_{17}$} & \SetScale{0.2}
\begin{picture}(90,48) (0,0)
\SetWidth{2.5}
\SetColor{Black}
\Vertex(120,45){6}
\Vertex(120,195){6}
\Vertex(330,45){6}
\Vertex(330,195){6}
\Line(120,195)(330,45)
\Line(330,195)(120,45)
\DashArrowLine(330,45)(330,195){10}
\Vertex(50,220){6}
\ArrowLine(50,220)(120,195)
\ArrowLine(0,240)(50,220)
\ArrowLine(330,195)(450,240)
\ArrowLine(450,0)(330,45)
\ArrowLine(120,45)(0,0)
\Vertex(120,130){6}
\DashArrowLine(120,195)(120,130){10}
\DashArrowLine(120,130)(120,45){10}
\GlueArc(137.5,215.83)(87.6,177.27,258.48){4}{7.71}
\end{picture} &
\raisebox{24pt}{$8$}
\\
[0.11cm] \raisebox{24pt}{$A_{18}$} & \SetScale{0.2}
\begin{picture}(90,48) (0,0)
\SetWidth{2.5}
\SetColor{Black}
\ArrowLine(330,195)(450,240)
\Line(120,195)(330,195)
\ArrowLine(450,0)(330,45)
\Line(330,45)(120,45)
\Vertex(120,45){6}
\Vertex(120,195){6}
\Vertex(330,45){6}
\Vertex(330,195){6}
\Vertex(50,220){6}
\ArrowLine(50,220)(120,195)
\ArrowLine(0,240)(50,220)
\ArrowLine(120,45)(0,0)
\DashArrowLine(120,195)(120,45){10}
\Vertex(330,120){6}
\DashArrowLine(330,45)(330,120){10}
\DashArrowLine(330,120)(330,195){10}
\GlueArc(257.38,358.66)(249.47,-146.23,-73.08){4}{21.43}
\end{picture} &
\raisebox{24pt}{$B_{18}$} & \SetScale{0.2}
\begin{picture}(90,48) (0,0)
\SetWidth{2.5}
\SetColor{Black}
\Vertex(120,45){6}
\Vertex(120,195){6}
\Vertex(330,45){6}
\Vertex(330,195){6}
\Line(120,195)(330,45)
\Line(330,195)(120,45)
\Vertex(50,220){6}
\ArrowLine(50,220)(120,195)
\ArrowLine(0,240)(50,220)
\ArrowLine(330,195)(450,240)
\ArrowLine(450,0)(330,45)
\ArrowLine(120,45)(0,0)
\DashArrowLine(120,195)(120,45){10}
\Vertex(330,135){6}
\DashArrowLine(330,45)(330,135){10}
\DashArrowLine(330,135)(330,195){10}
\GlueArc(121.72,-47.42)(276.87,41.21,105.01){4}{20.57}
\end{picture} &
\raisebox{24pt}{$8$}
\\
[0.11cm] \raisebox{24pt}{$A_{56}$} & \SetScale{0.2}
\begin{picture}(90,48) (0,0)
\SetWidth{2.5}
\SetColor{Black}
\Vertex(120,45){6}
\Vertex(120,195){6}
\Vertex(330,45){6}
\Vertex(330,195){6}
\ArrowLine(120,45)(0,0)
\ArrowLine(0,240)(120,195)
\ArrowLine(450,0)(330,45)
\ArrowLine(330,195)(450,240)
\Line(120,195)(330,195)
\Line(330,45)(120,45)
\DashArrowLine(330,45)(330,195){10}
\DashArrowLine(120,195)(120,45){10}
\Vertex(225,195){6}
\Vertex(225,45){6}
\Gluon(225,195)(225,45){4}{9.43}
\end{picture} &
\raisebox{24pt}{$B_{56}$} & \SetScale{0.2}
\begin{picture}(90,48) (0,0)
\SetWidth{2.5}
\SetColor{Black}
\Vertex(120,45){6}
\Vertex(120,195){6}
\Vertex(330,45){6}
\Vertex(330,195){6}
\ArrowLine(120,45)(0,0)
\ArrowLine(0,240)(120,195)
\ArrowLine(450,0)(330,45)
\ArrowLine(330,195)(450,240)
\Line(120,195)(330,45)
\Line(330,195)(120,45)
\DashArrowLine(330,45)(330,195){10}
\DashArrowLine(120,195)(120,45){10}
\Vertex(155,170){6}
\Vertex(295,170){6}
\GlueArc(225,57.5)(132.5,58.11,121.89){4}{9.43}
\end{picture} &
\raisebox{24pt}{$2$}
\\
[0.11cm] \raisebox{24pt}{$A_{57}$} & \SetScale{0.2}
\begin{picture}(90,48) (0,0)
\SetWidth{2.5}
\SetColor{Black}
\Vertex(120,45){6}
\Vertex(120,195){6}
\Vertex(330,45){6}
\Vertex(330,195){6}
\ArrowLine(120,45)(0,0)
\ArrowLine(0,240)(120,195)
\ArrowLine(450,0)(330,45)
\ArrowLine(330,195)(450,240)
\Line(120,195)(330,195)
\Line(330,45)(120,45)
\DashArrowLine(330,45)(330,195){10}
\Vertex(190,195){6}
\Vertex(120,130){6}
\DashArrowLine(120,195)(120,130){10}
\DashArrowLine(120,130)(120,45){10}
\GlueArc(111.84,208.98)(79.4,-84.1,-10.14){4}{6}
\end{picture} &
\raisebox{24pt}{$B_{57}$} & \SetScale{0.2}
\begin{picture}(90,48) (0,0)
\SetWidth{2.5}
\SetColor{Black}
\Vertex(120,45){6}
\Vertex(120,195){6}
\Vertex(330,45){6}
\Vertex(330,195){6}
\ArrowLine(120,45)(0,0)
\ArrowLine(0,240)(120,195)
\ArrowLine(450,0)(330,45)
\ArrowLine(330,195)(450,240)
\Line(120,195)(330,45)
\Line(330,195)(120,45)
\DashArrowLine(330,45)(330,195){10}
\Vertex(120,120){6}
\Vertex(180,150){6}
\DashArrowLine(120,195)(120,120){10}
\DashArrowLine(120,120)(120,45){10}
\GlueArc(134.58,165.83)(48.1,-107.65,-19.22){4}{4.29}
\end{picture} &
\raisebox{24pt}{$8$}
\\
[0.11cm] \raisebox{24pt}{$A_{78}$} & \SetScale{0.2}
\begin{picture}(90,48) (0,0)
\SetWidth{2.5}
\SetColor{Black}
\ArrowLine(330,195)(450,240)
\Line(120,195)(330,195)
\Line(330,45)(120,45)
\ArrowLine(120,45)(0,0)
\Vertex(120,45){6}
\Vertex(120,195){6}
\Vertex(330,45){6}
\Vertex(330,195){6}
\ArrowLine(0,240)(120,195)
\ArrowLine(450,0)(330,45)
\Vertex(120,120){6}
\Vertex(330,120){6}
\DashArrowLine(120,195)(120,120){10}
\DashArrowLine(120,120)(120,45){10}
\DashArrowLine(330,45)(330,120){10}
\DashArrowLine(330,120)(330,195){10}
\Gluon(120,120)(330,120){4}{13.71}
\end{picture} &
\raisebox{24pt}{$B_{78}$} & \SetScale{0.2}
\begin{picture}(90,48) (0,0)
\SetWidth{2.5}
\SetColor{Black}
\Vertex(120,45){6}
\Vertex(120,195){6}
\Vertex(330,45){6}
\Vertex(330,195){6}
\ArrowLine(120,45)(0,0)
\ArrowLine(0,240)(120,195)
\ArrowLine(450,0)(330,45)
\ArrowLine(330,195)(450,240)
\Line(120,195)(330,45)
\Line(330,195)(120,45)
\Vertex(120,120){6}
\DashArrowLine(120,195)(120,120){10}
\DashArrowLine(120,120)(120,45){10}
\Vertex(330,120){6}
\DashArrowLine(330,45)(330,120){10}
\DashArrowLine(330,120)(330,195){10}
\GlueArc(225,288.75)(198.75,-121.89,-58.11){4}{14.57}
\end{picture} &
\raisebox{24pt}{$2$}
\end{tabular*}
\caption{\small\sl NLO diagrams generated by gluon corrections to A-type
and B-type LO topologies.} \label{fig:diagsNLO-gluon}
\end{figure}
\begin{figure}[p]
\begin{tabular*}{14.2cm}{cc@{\hspace{1.5cm}}cc@{\hspace{1.5cm}}c}
{\bf A-type} & {\bf graph} & {\bf B-type} & {\bf graph} & {\bf \#}\\
[0.1cm]
\hline \hline
\\
\raisebox{24pt}{$A_{55g}$} & \SetScale{0.2}
\begin{picture}(90,48) (0,2) 
\SetWidth{2.5}
\SetColor{Black}
\Vertex(120,56){6}
\Vertex(120,206){6}
\Vertex(330,56){6}
\Vertex(330,206){6}
\ArrowLine(120,56)(0,11)
\ArrowLine(0,251)(120,206)
\ArrowLine(450,11)(330,56)
\ArrowLine(330,206)(450,251)
\Line(120,206)(330,206)
\Line(330,56)(120,56)
\DashArrowLine(330,56)(330,206){10}
\DashArrowLine(120,206)(120,56){10}
\Vertex(180,206){6}
\Vertex(275,206){6}
\GlueArc(227.5,203.5)(47.57,3.01,176.99){4}{8.57}
\end{picture} &
\raisebox{24pt}{$B_{55g}$} & \SetScale{0.2}
\begin{picture}(90,48) (0,0)
\SetWidth{2.5}
\SetColor{Black}
\Vertex(120,45){6}
\Vertex(120,195){6}
\Vertex(330,45){6}
\Vertex(330,195){6}
\ArrowLine(120,45)(0,0)
\ArrowLine(0,240)(120,195)
\ArrowLine(450,0)(330,45)
\ArrowLine(330,195)(450,240)
\Line(120,195)(330,45)
\Line(330,195)(120,45)
\DashArrowLine(330,45)(330,195){10}
\DashArrowLine(120,195)(120,45){10}
\Vertex(140,180){6}
\Vertex(205,135){6}
\GlueArc(167.5,150.28)(40.49,-22.17,132.78){4}{6}
\end{picture}
& \raisebox{24pt}{$4$}
\\
[0.11cm] \raisebox{24pt}{$A_{55q}$} & \SetScale{0.2}
\begin{picture}(90,48) (0,0)
\SetWidth{2.5}
\SetColor{Black}
\Vertex(120,45){6}
\Vertex(120,195){6}
\Vertex(330,45){6}
\Vertex(330,195){6}
\ArrowLine(120,45)(0,0)
\ArrowLine(0,240)(120,195)
\ArrowLine(450,0)(330,45)
\ArrowLine(330,195)(450,240)
\Line(330,45)(120,45)
\DashArrowLine(330,45)(330,195){10}
\DashArrowLine(120,195)(120,45){10}
\Vertex(180,195){6}
\Vertex(275,195){6}
\Line(120,195)(180,195)
\Line(275,195)(330,195)
\ArrowArc(227.5,172.5)(52.56,25.35,154.65)
\DashArrowArc(227.5,217.5)(52.56,-154.65,-25.35){10}
\end{picture} &
\raisebox{24pt}{$B_{55q}$} & \SetScale{0.2}
\begin{picture}(90,48) (0,0)
\SetWidth{2.5}
\SetColor{Black}
\ArrowLine(450,0)(330,45)
\DashArrowLine(330,195)(330,45){10}
\Vertex(120,45){6}
\Vertex(120,195){6}
\Vertex(330,45){6}
\Vertex(330,195){6}
\ArrowLine(120,45)(0,0)
\ArrowLine(0,240)(120,195)
\ArrowLine(330,195)(450,240)
\Line(330,195)(120,45)
\DashArrowLine(120,195)(120,45){10}
\Vertex(140,180){6}
\Vertex(205,135){6}
\Line(120,195)(140,180)
\Line(205,135)(330,45)
\ArrowArcn(157.94,136.47)(47.08,112.4,-1.79)
\DashArrowArcn(183.6,173.54)(44.08,299.04,171.57){10}
\end{picture} &
\raisebox{24pt}{$8 N_f$}
\\
[0.11cm] \raisebox{24pt}{$A_{77g}$} & \SetScale{0.2}
\begin{picture}(90,48) (0,0)
\SetWidth{2.5}
\SetColor{Black}
\Vertex(120,45){6}
\Vertex(120,195){6}
\Vertex(330,45){6}
\Vertex(330,195){6}
\ArrowLine(120,45)(0,0)
\ArrowLine(0,240)(120,195)
\ArrowLine(450,0)(330,45)
\ArrowLine(330,195)(450,240)
\Line(120,195)(330,195)
\Line(330,45)(120,45)
\DashArrowLine(330,45)(330,195){10}
\Vertex(120,155){6}
\Vertex(120,80){6}
\DashArrowLine(120,195)(120,155){10}
\DashArrowLine(120,155)(120,80){10}
\DashArrowLine(120,80)(120,45){10}
\GlueArc(117.5,117.5)(37.58,86.19,273.81){4}{7.71}
\end{picture} &
\raisebox{24pt}{$B_{77g}$} & \SetScale{0.2}
\begin{picture}(90,48) (0,0)
\SetWidth{2.5}
\SetColor{Black}
\Vertex(120,45){6}
\Vertex(120,195){6}
\Vertex(330,45){6}
\Vertex(330,195){6}
\ArrowLine(120,45)(0,0)
\ArrowLine(0,240)(120,195)
\ArrowLine(450,0)(330,45)
\ArrowLine(330,195)(450,240)
\Line(120,195)(330,45)
\Line(330,195)(120,45)
\DashArrowLine(330,45)(330,195){10}
\Vertex(120,155){6}
\Vertex(120,80){6}
\DashArrowLine(120,195)(120,155){10}
\DashArrowLine(120,80)(120,45){10}
\DashArrowLine(120,155)(120,80){10}
\GlueArc(117.5,117.5)(37.58,86.19,273.81){4}{7.71}
\end{picture} &
\raisebox{24pt}{$4$}
\\
[0.11cm] \raisebox{24pt}{$A_{77q}$} & \SetScale{0.2}
\begin{picture}(90,48) (0,0)
\SetWidth{2.5}
\SetColor{Black}
\Vertex(120,45){6}
\Vertex(120,195){6}
\Vertex(330,45){6}
\Vertex(330,195){6}
\ArrowLine(120,45)(0,0)
\ArrowLine(0,240)(120,195)
\ArrowLine(450,0)(330,45)
\ArrowLine(330,195)(450,240)
\Line(120,195)(330,195)
\Line(330,45)(120,45)
\DashArrowLine(330,45)(330,195){10}
\Vertex(120,155){6}
\Vertex(120,80){6}
\DashArrowLine(120,195)(120,155){10}
\DashArrowLine(120,80)(120,45){10}
\CArc(111.67,117.5)(38.41,-77.47,77.47)
\ArrowArc(128.33,117.5)(38.41,102.53,257.47)
\end{picture} &
\raisebox{24pt}{$B_{77q}$} & \SetScale{0.2}
\begin{picture}(90,48) (0,0)
\SetWidth{2.5}
\SetColor{Black}
\Vertex(120,45){6}
\Vertex(120,195){6}
\Vertex(330,45){6}
\Vertex(330,195){6}
\ArrowLine(120,45)(0,0)
\ArrowLine(0,240)(120,195)
\ArrowLine(450,0)(330,45)
\ArrowLine(330,195)(450,240)
\Line(120,195)(330,45)
\Line(330,195)(120,45)
\DashArrowLine(330,45)(330,195){10}
\Vertex(120,155){6}
\Vertex(120,80){6}
\DashArrowLine(120,195)(120,155){10}
\DashArrowLine(120,80)(120,45){10}
\ArrowArc(135.5,117.5)(40.58,112.46,247.54)
\CArc(104.5,117.5)(40.58,-67.54,67.54)
\end{picture} &
\raisebox{24pt}{$4$}
\\
[0.11cm] \raisebox{24pt}{$A_{77s}$} & \SetScale{0.2}
\begin{picture}(90,48) (0,0)
\SetWidth{2.5}
\SetColor{Black}
\Vertex(120,45){6}
\Vertex(120,195){6}
\Vertex(330,45){6}
\Vertex(330,195){6}
\ArrowLine(120,45)(0,0)
\ArrowLine(0,240)(120,195)
\ArrowLine(450,0)(330,45)
\ArrowLine(330,195)(450,240)
\Line(120,195)(330,195)
\Line(330,45)(120,45)
\DashArrowLine(330,45)(330,195){10}
\Vertex(120,120){6}
\DashArrowLine(120,195)(120,120){10}
\DashArrowLine(120,120)(120,45){10}
\DashArrowArc(85,120)(35,0,360){10}
\end{picture} &
\raisebox{24pt}{$B_{77s}$} & \SetScale{0.2}
\begin{picture}(90,48) (0,0)
\SetWidth{2.5}
\SetColor{Black}
\Vertex(120,45){6}
\Vertex(120,195){6}
\Vertex(330,45){6}
\Vertex(330,195){6}
\ArrowLine(120,45)(0,0)
\ArrowLine(0,240)(120,195)
\ArrowLine(450,0)(330,45)
\ArrowLine(330,195)(450,240)
\Line(120,195)(330,45)
\Line(330,195)(120,45)
\DashArrowLine(330,45)(330,195){10}
\Vertex(120,120){6}
\DashArrowLine(120,195)(120,120){10}
\DashArrowLine(120,120)(120,45){10}
\DashArrowArc(85,120)(35,0,360){10}
\end{picture} &
\raisebox{24pt}{$4$}
\end{tabular*}
\caption{\small\sl NLO diagrams generated by self-energy corrections to
A-type and B-type LO topologies.} \label{fig:diagsNLO-self}
\end{figure}
\begin{figure}[p]
\begin{tabular*}{14.2cm}{cc@{\hspace{1.5cm}}cc@{\hspace{1.5cm}}c}
{\bf A-type} & {\bf graph} & {\bf B-type} & {\bf graph} & {\bf \#}\\
[0.1cm]
\hline \hline
\\
\raisebox{24pt}{$A_{V}$} & \SetScale{0.2}
\begin{picture}(90,48) (0,0) 
\SetWidth{2.5}
\SetColor{Black}
\Vertex(120,45){6}
\Vertex(120,195){6}
\Vertex(330,45){6}
\Vertex(330,195){6}
\ArrowLine(120,45)(0,0)
\ArrowLine(0,240)(120,195)
\ArrowLine(450,0)(330,45)
\ArrowLine(330,195)(450,240)
\Line(330,45)(120,45)
\DashArrowLine(330,45)(330,195){10}
\Vertex(120,120){6}
\Vertex(210,195){6}
\DashArrowLine(120,120)(120,45){10}
\Line(120,195)(120,120)
\Line(210,195)(330,195)
\DashArrowLine(120,195)(210,195){10}
\ArrowArcn(114.17,218.5)(98.67,-13.78,-86.61)
\end{picture} &
\raisebox{24pt}{$B_{V}$} & \SetScale{0.2}
\begin{picture}(90,48) (0,0)
\SetWidth{2.5}
\SetColor{Black}
\Vertex(120,45){6}
\Vertex(120,195){6}
\Vertex(330,45){6}
\Vertex(330,195){6}
\ArrowLine(120,45)(0,0)
\ArrowLine(0,240)(120,195)
\ArrowLine(450,0)(330,45)
\ArrowLine(330,195)(450,240)
\DashArrowLine(330,45)(330,195){10}
\Vertex(120,120){6}
\DashArrowLine(120,120)(120,45){10}
\Line(120,195)(120,120)
\Line(120,45)(330,195)
\Vertex(175,155){6}
\Line(175,155)(330,45)
\DashArrowLine(120,195)(175,155){10}
\ArrowArcn(130,165)(46.1,-12.53,-102.53)
\end{picture}
& \raisebox{24pt}{$4$}
\\
[0.11cm] \raisebox{24pt}{$A_{T}$} & \SetScale{0.2}
\begin{picture}(90,48) (0,0)
\SetWidth{2.5}
\SetColor{Black}
\Vertex(120,45){6}
\Vertex(120,195){6}
\Vertex(330,45){6}
\Vertex(330,195){6}
\ArrowLine(120,45)(0,0)
\ArrowLine(0,240)(120,195)
\ArrowLine(450,0)(330,45)
\ArrowLine(330,195)(450,240)
\DashArrowLine(120,195)(120,45){10}
\Line(120,195)(225,195)
\Vertex(225,195){6}
\Vertex(225,45){6}
\Line(225,45)(120,45)
\ArrowLine(225,45)(225,195)
\DashArrowLine(225,195)(330,195){10}
\DashArrowLine(330,45)(225,45){10}
\Line(330,45)(330,195)
\end{picture} &
\raisebox{24pt}{$B_{T}$} & \SetScale{0.2}
\begin{picture}(90,48) (0,0)
\SetWidth{2.5}
\SetColor{Black}
\Vertex(120,45){6}
\Vertex(120,195){6}
\Vertex(330,45){6}
\Vertex(330,195){6}
\ArrowLine(120,45)(0,0)
\ArrowLine(0,240)(120,195)
\ArrowLine(450,0)(330,45)
\ArrowLine(330,195)(450,240)
\DashArrowLine(120,195)(120,45){10}
\Vertex(290,75){6}
\Vertex(290,165){6}
\ArrowLine(290,75)(290,165)
\Line(290,165)(120,45)
\Line(120,195)(290,75)
\DashArrowLine(330,45)(290,75){10}
\DashArrowLine(290,165)(330,195){10}
\Line(330,195)(330,45)
\end{picture} &
\raisebox{24pt}{$16$}
\end{tabular*}
\caption{\small\sl NLO diagrams generated by squark corrections to A-type
and B-type LO topologies.} \label{fig:diagsNLO-TV}
\end{figure}
\begin{figure}[t!!!]
\begin{center}
\begin{tabular*}{8.1cm}{cc@{\hspace{1.5cm}}c}
{\bf diagram} & {\bf graph}  & {\bf \#}\\
[0.1cm]
\hline \hline
\\
\raisebox{24pt}{$X_{1}$} &
\SetScale{0.2}
\begin{picture}(90,48) (0,0) 
\SetWidth{2.5}
\SetColor{Black}
\Line(120,195)(120,45)
\ArrowLine(120,45)(0,0)
\ArrowLine(450,0)(330,45)
\Line(330,45)(330,195)
\Vertex(120,45){6}
\Vertex(120,195){6}
\Vertex(330,45){6}
\Vertex(330,195){6}
\ArrowLine(0,240)(120,195)
\ArrowLine(330,195)(450,240)
\Vertex(225,120){6}
\DashArrowLine(120,195)(225,120){10}
\DashArrowLine(225,120)(330,195){10}
\DashArrowLine(225,120)(120,45){10}
\DashArrowLine(330,45)(225,120){10}
\end{picture}
& \raisebox{24pt}{$2$}
\\
[0.11cm]
\raisebox{24pt}{$X_{2}$} &
\SetScale{0.2}
\begin{picture}(90,48) (0,0)
\SetWidth{2.5}
\SetColor{Black}
\ArrowLine(120,45)(0,0)
\Vertex(120,45){6}
\Vertex(120,195){6}
\Vertex(330,45){6}
\Vertex(330,195){6}
\ArrowLine(0,240)(120,195)
\Vertex(225,120){6}
\DashArrowLine(120,195)(225,120){10}
\DashArrowLine(225,120)(330,195){10}
\DashArrowLine(225,120)(120,45){10}
\DashArrowLine(330,45)(225,120){10}
\Line(120,195)(330,195)
\Line(330,45)(120,45)
\ArrowLine(450,0)(330,195)
\ArrowLine(330,45)(450,240)
\end{picture} &
\raisebox{24pt}{$1$}
\end{tabular*}
\end{center}
\caption{\small\sl NLO diagrams generated by four squark interaction
vertices.}
\label{fig:diagsNLO-X}
\end{figure}
They have been generated by using the {\Math} \cite{math} package {\FA}
\cite{FA}. The full set of NLO diagrams can be divided in four categories.
\begin{enumerate}
\item Gluon corrections, connecting different legs in A-type or
B-type LO diagrams. These corrections are collected in
fig.~\ref{fig:diagsNLO-gluon}.
\item Self-energy corrections of internal legs; these are collected in
fig.~\ref{fig:diagsNLO-self}.
\item Squark corrections, generated by adding to the LO topologies one
more squark propagator via the quark-squark-gluino interaction vertex. These
diagrams are shown in fig.~\ref{fig:diagsNLO-TV}.
\item Quartic scalar interactions, generated by the four squark vertex
and collected in fig.~\ref{fig:diagsNLO-X}.
\end{enumerate}
All these diagrams, except for those belonging to the last category, are
ge\-ne\-ra\-ted from the LO topologies with the inclusion of an additional
loop in all possible ways. The last column in figs.
\ref{fig:diagsNLO-gluon}-\ref{fig:diagsNLO-X} indicates the number of
existing diagrams, including the one shown in the figure, that are obtained
from the latter by performing $90^{\rm o}$ or $180^{\rm o}$ rotations around
the horizontal, vertical or perpendicular axis. Diagrams containing
self-energy corrections of the external legs have not been included in the
above list. As discussed in section~\ref{sec:EffHam}, however, these
corrections have to be taken into account and receive two kinds of
contributions. QCD contributions mediated by gluons enter the calculation of
both the full and the effective theory and cancel in the matching, while
supersymmetric squark and gluino corrections give a finite contribution to
the NLO Wilson coefficients.

Among the diagrams presented in
figs.~\ref{fig:diagsNLO-gluon}-\ref{fig:diagsNLO-X}, those producing either
UV or IR divergences are the following ones,
\beqn
&\mbox{UV divergent:} &
\{A_{15},A_{17},A_{57},A_{55g},A_{55q},A_{77g},A_{77q},A_{77s}\} +
\{A \ra B\} \nn \\
&\mbox{IR divergent:} &
\{A_{12},A_{13},A_{14}\}+\{A \ra B\}~.
\label{UVIRdiags}
\eeqn
By looking at figs.~\ref{fig:diagsNLO-gluon}-\ref{fig:diagsNLO-X} one can
see that UV divergent graphs are only those containing vertex and
self-energy corrections. These graphs provide in particular the SUSY
contributions to the renormalization of the strong coupling constant and of
the squark and gluino fields and masses. IR divergences, instead, are
produced by those diagrams in which a virtual gluon connects two external
quark lines. These diagrams are in a one-to-one correspondence with the
diagrams entering the calculation in the effective theory and the whole set
of IR divergences cancel in the matching.

We now describe, in some detail, the procedure followed in the evaluation
of the two-loop diagrams of the full theory.

Having chosen external quarks with zero masses and momenta, a typical
two-loop amplitude can be schematically expressed as
\beq
\mc{D}=\int \frac{d^d q_1}{(2 \pi)^d} \frac{d^d q_2}{(2 \pi)^d}
\frac{\Gamma_A(q_1,q_2,\mu,\nu,\ldots)\otimes\
\Gamma_B(q_1,q_2,\mu,\nu,\ldots)}
{\left(q_1^2 - m_1^2\right)^{n_1} \left(q_2^2 - m_2^2\right)^{n_2}
\left((q_1-q_2)^2 - m_3^2 \right)^{n_3}}
\label{Dia2loop}
\eeq
where $\Gamma_{A,B}$ represent strings of gamma matrices and loop momenta
with sa\-tu\-ra\-ted Lorentz indices. To simplify the notation, external
quark spinors in the amplitude have been omitted. In the denominator,
partial fractioning has been applied in order to express it in terms of the
minimum number of scalar propagators, which is equal to three for a two-loop
calculation with vanishing external momenta. The masses $m_{1,2,3}$ stand
generically for the different masses entering the calculation, namely the
gluino mass $M_{\tilde{g}}$, the mean squark mass $M_s$ defined in
eq.~(\ref{app:miaeq}) and, when regularizing with a massive gluon, the gluon
mass $\lambda$.

One of the advantages of working with vanishing external momenta is that,
once the loop integration has been performed, the amplitude in
eq.~(\ref{Dia2loop}) turns out to be expressed only in terms of strings of
gamma matrices, with either physical ($\ov{\Ga}_A^{(i)} \otimes
\ov{\Ga}_B^{(i)}$) or evanescent ($E_A^{(i)} \otimes E_B^{(i)}$) structures,
multiplied by scalar functions of the particle masses:
\beq
\mc{D}= \sum_i \left[ a_i(m) \ \ov{\Ga}_A^{(i)} \otimes \ov{\Ga}_B^{(i)}~ +
~ b_i(m)\ E_A^{(i)} \otimes E_B^{(i)} \right]
\label{Dia2loop-2}
\eeq
The functions $b_i(m)$ are not of interest for our purposes, since the
evaluation of the Wilson coefficients at the NLO only requires, according to
eq.~(\ref{match-passage}), the projections $F_i^{(1)}$ of the two-loop
amplitude on the physical operators. The complete basis of Lorentz invariant
Dirac structures on which we project is given by
\beq
\ov{\Ga}_A^{(i)} \otimes \ov{\Ga}_B^{(i)}
= \{ \ga^\mu_L \otimes \ga_{\mu L},~ \ga^\mu_L \otimes \ga_{\mu R},~
P_L \otimes P_L, ~ P_L \otimes P_R, ~
\sigma^{\mu \nu}_L \otimes \sigma_{\mu \nu L}\}+ \{ L\leftrightarrow R \}
\label{projector-structures}
\eeq
where $L\leftrightarrow R$ indicates the structures obtained by exchanging
left and right projectors.

In order to extract directly from a given amplitude $\mc{D}$ the
coefficients $a_i$ of the physical operators, we used a basis of orthonormal
projectors. These are defined as a set of strings of gamma matrices,
$P_A^{(j)} \otimes P_B^{(j)}$, satisfying the orthonormality conditions
\beq
{\rm Tr}\left[\ov{\Ga}_A^{(i)}\, P_A^{(j)}\, \ov{\Ga}_B^{(i)}\, P_B^{(j)}
\right] = \de_{ij} ~.
\label{projectors}
\eeq
In the DRED scheme the traces are computed in four dimensions. In NDR
instead, where gamma matrices are $d$-dimensional objects, the traces are
performed in $d$ dimensions and the orthonormality conditions
(\ref{projectors}) are required to be fulfilled up to and including terms of
${\cal O}(\vep)$; this is sufficient, since the two-loop amplitude in the
present calculation contains at most $1/\vep$ divergences. With these
requirements the projectors $P_A^{(j)} \otimes P_B^{(j)}$ are uniquely
defined. The main advantage of using this procedure is that, once the
projection is applied to an amplitude of the form (\ref{Dia2loop}), the
resulting expression only involves scalar integrals. The number of
independent two-loop integrations to be performed is therefore significantly
reduced.

Besides satisfying eq.~(\ref{projectors}), the projectors must be also
orthogonal to the evanescent structures. This requirement ensures that, once
the projection is applied to the r.h.s. of eq.~(\ref{Dia2loop-2}), no finite
contribution coming from the evanescent operators is kept in the amplitude.
This issue is of relevance in the DRED-$d$ scheme, where IR divergences are
dimensionally regularized. In this case, the orthogonality of the projectors
to the evanescent operators is guaranteed by the following observation: all
the Dirac structures entering the evanescent operators in this scheme have
uncontracted Lorentz indices and, after the four dimensional projections,
can only give rise to products of four dimensional $\tilg_{\mu\nu}$ tensors.
The latter, in turn, are orthogonal to the evanescent operators in the DRED
scheme defined as in eq.~(\ref{basis-ev}), because of the second of
eqs.~(\ref{dred}).

After the projection has been performed, the evaluation of the two-loop
integrals is reduced to computing scalar integrals of the form
\beqn
&&\mc{I}(m_1,m_2,m_3;n_1,n_2,n_3) \equiv \nn \\
&&\int \frac{d^d q_1}{(2 \pi)^d} \frac{d^d q_2}{(2 \pi)^d}
\frac{1}{\left(q_1^2 - m_1^2\right)^{n_1} \left(q_2^2 -
m_2^2\right)^{n_2} \left( (q_1-q_2)^2 - m_3^2 \right)^{n_3}}~.
\label{Ilan1n2n3}
\eeqn
This task is greatly simplified by the use of the recurrence
relations~\cite{Chetyrkin:1981qh}, which allow to reduce all scalar
integrals of the form~(\ref{Ilan1n2n3}) to a single two-loop master
integral, $\mc{I}(m_1,m_2,m_3;1,1,1)$, besides trivial one-loop tadpole
integrals.\footnote{The application of recurrence relations can be
automatically performed by using the Tarasov reduction
algorithm~\cite{tarasov-th,tarasov-ph} implemented in the {\Math} program
{\sf TARCER}~\cite{TARCER}.} The result for the master integral
$\mc{I}(m_1,m_2,m_3;1,1,1)$ is given in ref.~\cite{Davydychev:1992mt}.

A further step is required when one of the three masses in the denominator
of the integral~(\ref{Ilan1n2n3}) is the gluon mass $\lambda$, introduced to
regularize IR divergences. As a result of having implemented the recurrence
relations, one finds that the coefficients multiplying the master integral
contain negative powers of $\lambda$, up to $\mc{O}(1/\lambda^4)$. The
master integral itself must be therefore expanded up to $\mc{O}(\lambda^4)$.
After the expansion, all power divergences must cancel in the amplitude
and only logarithmic IR divergences remain, which cancel in the matching.

The last step, after the projection and the loop integration, consists in
expressing the NLO amplitude in terms of tree-level matrix elements of the
operators in the basis~(\ref{basis}). This is done by using Fierz
rearrangement and color algebra. Note, however, that the possibility of
expressing the amplitude in terms of tree-level matrix elements, up to terms
of $\mc{O}(\vep)$, does not occur diagram by diagram. It only holds, in
general, for the complete amplitude. This step already provides, therefore,
a useful check of the correctness of the calculation.

The sum of the UV renormalized and IR regularized NLO diagrams gives, in the
notation of eq.~(\ref{match-passage}), the functions $F_j^{(1)}$, that
represent the main ingredient in the NLO evaluation of the Wilson
coefficients.

\section{Calculation in the effective theory\label{sec:NLOeff}}
\setcounter{equation}{0}
The second step required in the matching procedure is the calculation
of the amplitude in the effective theory and, in particular, of the matrix
$r$ defined in eq.~(\ref{Aeff-form}). Using this equation and introducing
the renormalization matrix $Z$ for the operators $\mc{O}_i$, we can write
the one-loop matrix elements of the renormalized operators as
\beq
\< \mc{O}_i \>^{ren} = \sum_j Z^{-1}_{ij}\, \< \mc{O}_j \>^{bare}
= \sum_j \left( 1+ \as \,r \right)_{ij}\< \mc{O}_j \>^{(0)}~.
\label{r-def}
\eeq
We note again that, in the case of the DRED-$d$ regularization setup, the
first index $i$ of $r_{ij}$ runs over the evanescent operators too. The
reason is that in the presence of dimensionally regularized IR divergences
the renormalized matrix elements of evanescent operators do not vanish.

Eq.~(\ref{r-def}) shows that the calculation of the matrix $r$ involves two
steps: i) the determination of the matrix elements of the bare operators $\<
\mc{O}_j \>^{bare}$ up to one loop and ii) the one loop determination of the
renormalization matrix $Z$.

\begin{figure}[t]
\begin{center}
\includegraphics[scale=0.4]{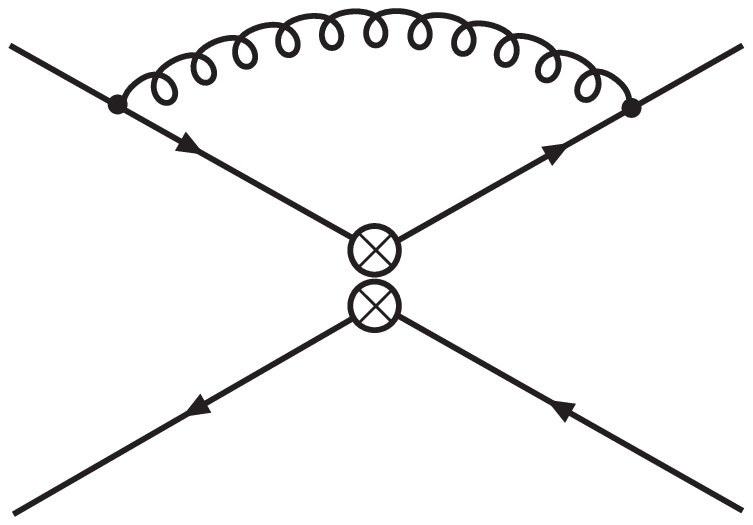}
\hspace{0.15cm} \includegraphics[scale=0.4]{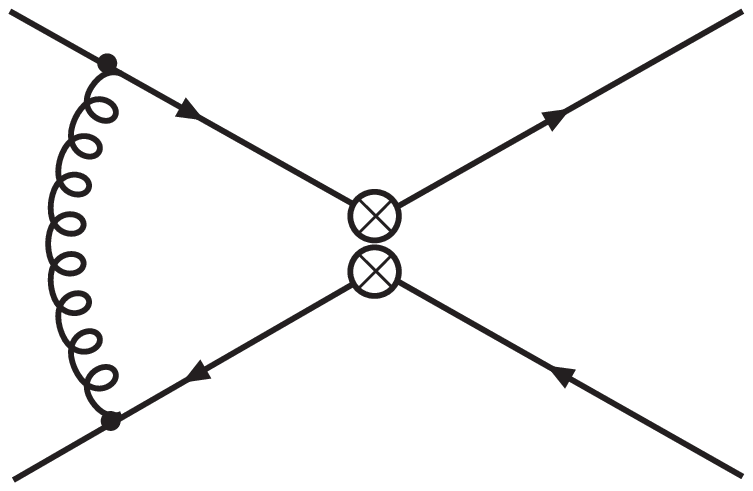}
\hspace{0.15cm} \includegraphics[scale=0.4]{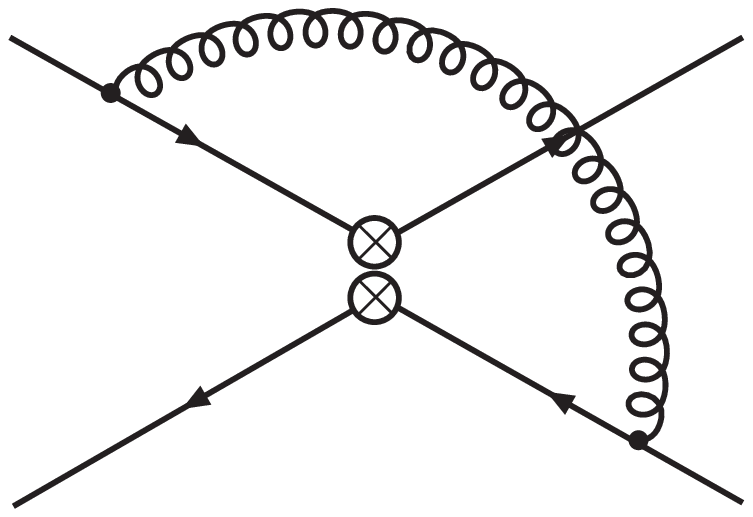}
\caption{\small\sl Feynman diagrams contributing at one loop to the
four-fermion operator matrix elements in the effective theory.} 
\label{fig:diags-eff}
\end{center}
\end{figure}

As for the calculation of the bare matrix elements, they receive
contributions in the effective theory only from QCD interactions. The
relevant Feynman diagrams are those represented in fig.~\ref{fig:diags-eff},
plus the three diagrams obtained by performing $180^{\rm o}$ rotations.
Consistency in the matching procedure requires the matrix elements in the
effective theory to be computed between the same set of external states and
with the same regularization procedure for IR divergences adopted in the
full theory. Therefore, we have performed this calculation by choosing
massless quarks with zero momentum as external states and implementing
separately the three regularization setups: DRED-$d$, DRED-$\la$ and
NDR-$\la$. Note, in particular, that the bare amplitudes vanish identically
at one loop in the DRED-$d$ scheme, since all loop integrals in this case
reduce to tadpole massless integrals which vanish in dimensional
regularization.

Eq.~(\ref{r-def}) also indicates that the one loop results for the bare
matrix elements must be projected onto the basis of the physical operators.
This projection implies a definition of the evanescent operators. In the
DRED regularization scheme the only evanescent operators entering the
calculation are defined to be proportional to the tensor $\Delta g_{\mu\nu}$
of eq.~(\ref{app:gsplit}). Besides the operators specified in
eq.~(\ref{basis-ev}), we also find the appearance of the evanescent
operators
\beqn
E^{\rm DRED}_4 &=& \Delta g_{\mu\nu}\;
 \bar{d}^i {\sigma}^{\mu\rho}_L b^i\, \bar{d}^j {\sigma}^{\nu}_{\rho\, L}
b^j ~, \nn \\
E^{\rm DRED}_5 &=& \Delta g_{\mu\nu}\;
 \bar{d}^i {\sigma}^{\mu\rho}_L b^j\, \bar{d}^j {\sigma}^{\nu}_{\rho\, L}
b^i ~.
\label{ev45}
\eeqn

In the NDR scheme, instead, both Dirac and Fierz evanescent operators must
be introduced. Dirac evanescent operators are defined from the orthogonality
condition to the Dirac projectors (see eq.~(\ref{projectors})),
\beq
{\rm Tr}\left[E_A^{(i)}\, P_A^{(j)}\, E_B^{(i)}\, P_B^{(j)} \right] =0~.
\label{NDR-projectors}
\eeq 
The complete list is given in ref.~\cite{NLOADMcheck}. As for the Fierz
evanescent operators, they are defined without introducing in the four
dimensional Fierz relations arbitrary terms of $\mc{O}(\vep)$; for example,
the $\gamma_{\mu L}\otimes \gamma^{\mu}_L$ Fierz evanescent operator reads
\beq
E_1^{\rm NDR} = \dd^{i} \gamma_{\mu L} b^{j}\, \dd^{j} \gamma^{\mu}_L b^{i}
- \dd^{i} \gamma_{\mu L} b^{i}\, \dd^{j} \gamma^{\mu}_L b^{j} 
\eeq
and similarly for the other gamma structures.

According to eq.~(\ref{r-def}), the second ingredient in the determination
of the matrix $r$ is the one-loop calculation of the renormalization matrix
$Z$. This requires again the evaluation of the Feynman diagrams shown in
fig.~\ref{fig:diags-eff}. In this case however, in order to identify the UV
divergences within dimensional regularization, one can either regularize the
IR divergences with a fictitious gluon mass or consider a set of IR finite
external states, for instance off-shell quarks with fixed momentum $p$.

In both the $\MSbar$-DRED and -NDR regularization schemes the
renormalization matrix of the physical operators is determined by applying
the modified minimal subtraction prescription. Evanescent operators, 
instead, must satisfy a different renormalization condition. For IR finite
configurations of external states, this condition reads
\beqn
\<E_i(\mu)\> = 0 \quad \textrm{ in the limit}\; d\rightarrow 4~,
\label{Ei-condition} 
\eeqn
and holds at any value of the renormalization scale
$\mu$~\cite{Dugan:1990df}. It guarantees that the evanescent operators do
not play any role when going back to four dimensions and can be eventually
removed from the operator basis of the effective Hamiltonian.

The final result for the matrix $r$ defined in eq.~(\ref{r-def}) depends on
several choices done in the calculation: the external states, the IR
regulator (when IR divergences are present) and the renormalization scheme
of the local operators. Thus we end up with three different matrices,
$r^{{\rm DRED}-d}$, $r^{{\rm DRED}-\lambda}$ and $r^{{\rm NDR}-\lambda}$.
Here we only present the results for the differences $\Delta r$ between
these matrices because, at variance with the $r$'s, they are independent of
the specific choice of both the external states and the IR regulator. As
can be seen from eq.~(\ref{r-def}), the matrices $\Delta r$ provide the
relation between operators renormalized in different schemes. In the case of
the $\MSbar$-NDR and DRED schemes, for instance, this relation reads
\beq
\< {\cal O}_i \>^{\rm \msbar-NDR} = \left( 1 + \as \Delta r^{\rm NDR/DRED}
\right)_{ij} \< {\cal O}_j \>^{\rm \msbar-DRED}~,
\label{QiNDR-DRED}
\eeq
where $\Delta r^{\rm NDR/DRED} \equiv r^{\rm NDR}-r^{\rm DRED}$. For this
matrix we obtain the result
\beq
\Delta r^{\rm NDR/DRED}= \left(
\begin{array}{ccccc}
-3    & 0     & 0    & 0     & 0 \\
 0    & -13/3 & -1/3 & 0     & 0 \\
 0    & -29/6 &  7/6 & 0     & 0 \\
 0    & 0     & 0    & -5/3  & -3 \\
 0    & 0     & 0    & -7/2  & -1/6 \\
\end{array}
\right)~,
\label{DZnd} 
\eeq
in the basis ${\cal O}_1,\ldots,{\cal O}_5$ of eq.~(\ref{basis}). Since
chirality is conserved by QCD interactions in the limit of massless quarks,
the corresponding matrix for the operators $\tilde{{\cal O}}_{1,2,3}$ is
equal to the $3 \times 3$ submatrix for ${\cal O}_{1,2,3}$ in
eq.~(\ref{DZnd}) and the two sets of operators do not mix.

In addition, we provide the matrix $\Delta r$ relating the $\MSbar$-DRED
with the so called RI-MOM scheme in the Landau gauge~\cite{RI-MOM}. This is
useful because this scheme is frequently used in lattice QCD calculations of
the hadronic matrix elements. This matrix reads:
\beq
\Delta r^{\rm DRED/RI} = \left(
\begin{array}{ccc}
-\frac{5}{3} + 8\,\ln 2 & 0 & 0 \\
[0.2cm]
0 & A_{2 \times 2} &  0 \\
[0.2cm]
0 & 0 & B_{2 \times 2}
\end{array}
\right)~,
\label{Dr-dredri}
\eeq
with
\beq
A_{2 \times 2} = \left(
\begin{array}{cc}
\frac{67}{9} + \frac{44}{9}\,\ln 2 &
-\frac{1}{9} + \frac{28}{9}\,\ln 2 \\
[0.2cm]
-\frac{28}{9} + \frac{28}{9}\,\ln 2 &
-\frac{68}{9} + \frac{44}{9}\,\ln 2
\end{array}
\right)
\eeq
and
\beq
B_{2 \times 2} = \left(
\begin{array}{cc}
13-\frac{2}{3}\,\ln 2 &
1 + 2\,\ln 2 \\
[0.2cm]
\frac{11}{2} + 2\,\ln 2 &
- \frac{1}{2} -\frac{2}{3}\,\ln 2
\end{array} \right)~.
\eeq
The results in eqs.~(\ref{DZnd}) and (\ref{Dr-dredri}) can be also combined
to obtain the matrix relating the $\MSbar$-NDR with the RI-MOM scheme:
$\Delta r^{\rm NDR/RI} = \Delta r^{\rm NDR/DRED} + \Delta r^{\rm DRED/RI}$.

\section{Results and checks of the calculation\label{sec:NLOmatch}}
\setcounter{equation}{0} 
In the previous sections we have described the calculation of the two
ingredients needed to obtain the Wilson coefficients at the NLO: the LO and
NLO amplitudes in the full theory, $F^{(0)}$ and $F^{(1)}$, and the matrix
$r$ in the effective theory. The NLO Wilson coefficients are finally
determined using eq.~(\ref{match-passage}). They bear a dependence on both
the renormalization scheme and scale. These dependences only arise at the
NLO and allow one to perform useful checks of the calculation. The relations
among the results for the coefficients as obtained in the three
regularization setups, DRED-$\la$, NDR-$\la$ and DRED-$d$, will be discussed
in the following subsection. The scale dependence of the Wilson coefficients
must satisfy the renormalization group equation, and this constraint will be
addressed in subsection~\ref{sec:RGE}.

\subsection{Regularization and renormalization scheme dependence
\label{sec:schemes}}
The results for the coefficients obtained in the DRED-$\la$ setup must be
equal to those obtained in DRED-$d$, since the Wilson coefficients cannot
depend on the IR regulator. Indeed, upon explicit comparison, they are found
to be in agreement. We emphasize that this is a non-trivial check of the
calculation. Indeed, whereas the computation in the DRED-$\la$ scheme
presents basically no subtlety, the one in the DRED-$d$ regularization
entails the inclusion in the full theory of the LO contributions up to
$O(\vep)$ and of the evanescent operators. All these contributions should
sum up to reconstruct the results obtained by using the gluon mass as IR
regulator.

The results for the Wilson coefficients obtained in the $\MSbar$-DRED and
NDR renormalization schemes differ because the coefficients are scheme
dependent quantities. They can be compared using the scheme independence of
the effective Hamiltonian:
\beq
\<\ov{b},d|\mc{H}_{\rm eff}|b,\ov{d}\> = \sum_i C_i^{\rm DRED}
\<\ov{b},d|Q_i|b,\ov{d}\>^{\rm DRED} = C_i^{\rm NDR}
\<\ov{b},d|Q_i|b,\ov{d}\>^{\rm NDR}~.
\label{scheme-ch}
\eeq
The relation between renormalized operators in the two schemes has been
written in eq.~(\ref{QiNDR-DRED}) in term of the matrix $\Delta r$. From
eq.~(\ref{scheme-ch}), it then follows that the same matrix also relates the
coefficients in different schemes, \eg
\beq
C_i^{\rm NDR}(M_s,M_{\tilde{g}},\alpha_s) = \sum_j \left( 1 - \as \Delta
r^{\rm NDR/DRED}\right)_{ji} C_j^{\rm DRED}(M_s,M_{\tilde{g}},\alpha_s) ~,
\label{NDRtoDRED}
\eeq
where $\Delta r^{\rm NDR/DRED}$ is given in eq.~(\ref{DZnd}). Notice that
in eq.~(\ref{NDRtoDRED}) the transposed matrix $\Delta r^T$ enters. The
coupling constant $\alpha_s$ and the SUSY masses $M_s$ and $M_{\tilde{g}}$
in the previous equation are also scheme dependent quantities. This
dependence starts at $O(\alpha_s)$ and must be taken into account in the
matching at the NLO. In order to verify eq.~(\ref{NDRtoDRED}), therefore,
one needs to express all the couplings in the same scheme. The required
relations are~\cite{MV}:
\beqn
&& \hat{\alpha}_s^{\rm NDR} = \alpha_s^{\rm DRED} \left( 1 +
\frac{\alpha_s}{4\pi} (N_c - C_F) \right) \nn \\
&& M^{\rm NDR}_{\tilde{g}} = M^{\rm DRED}_{\tilde{g}} \left( 1 + \as N_c
\right) \label{MV-shifts} \\
&& M^{\rm NDR}_s = M^{\rm DRED}_s \left( 1 + \mc{O}(\alpha_s^2) \right) \nn
~,
\eeqn
where $N_c=3$ and $C_F=4/3$ are the SU(3$)_c$ color factors. The strong
coupling constant $\hat{\alpha}_s$ in eq.~(\ref{MV-shifts}) indicates the
coupling of the quark-squark-gluino vertex. A different relation is found
for the quark-quark-gluon coupling, which differs from $\hat{\alpha}_s$ in
the NDR scheme because this regularization breaks supersymmetry~\cite{MV}
(see eq.~(\ref{asdred})). In the present calculation, since the shifts
expressed by eq.~(\ref{MV-shifts}) are ${\cal O}(\alpha_s)$, they have to
be implemented only in the LO amplitude, where only the coupling
$\hat{\alpha}_s$ appears. We find that our results for the Wilson
coefficients as obtained in the DRED and NDR schemes consistently satisfy
eq.~(\ref{NDRtoDRED}), with the matrix $\Delta r^{\rm NDR/DRED}$ given in
eq.~(\ref{DZnd}).

\subsection{Renormalization scale dependence \label{sec:RGE}}
Beyond LO, the Wilson coefficients acquire an explicit dependence on the
renormalization scale $\mu$. This dependence is controlled by the
renormalization group equation, which provides therefore an additional check
of the calculation.

The renormalization group equation for the Wilson coefficients of the
MSSM \cite{Bertolini:1990if,Barbieri:1995tw} can be written as
\beqn
&& \left[\frac{\partial}{\partial \ln \mu^2}
+ \frac{d\alpha_s}{d \ln \mu^2} \frac{\partial}{\partial\alpha_s}
+ \frac{d M_{\tilde{g}}^2}{d \ln \mu^2} \frac{\partial}{\partial
M_{\tilde{g}}^2}
+ \frac{d M_s^2}{d \ln \mu^2} \frac{\partial}{\partial M_s^2}
+ \sum_X \frac{d \Delta_{X}}{d \ln \mu^2} \frac{\partial}{\partial
\Delta_{X}} \right. \nn \\
&& \qquad \left.
 - \frac{1}{2} \gamma^T \right] \vec{C}(\mu) =0~,
\label{chain-mu-dep}
\eeqn
and takes into account the scale dependence of all the quantities entering
the coefficients, namely the strong coupling constant $\alpha_s$, the
squark and gluino masses $M_s$ and $M_{\tilde{g}}$ and the dimensionful
mass insertions $\Delta_{X}$, with $X = LL,\,RR,\,LR\, ,RL$.

The matrix $\gamma$ in eq.~(\ref{chain-mu-dep}) is the anomalous dimension
matrix of the four-fermion operators~(\ref{basis}) in the effective theory
(\ie QCD). It can be expanded as
\beq
\ga(\alpha_s) = Z^{-1} \frac{d Z}{d \ln \mu} = \as \, \gamma_0 +
\mc{O}(\alpha_s^2)
\label{ADM-def}
\eeq
where, for the LO anomalous dimension $\gamma_0$, we obtain the expression
\beq
\gamma_0 =
\left(
\begin{array}{ccccc}
4 & 0     & 0    & 0   & 0 \\
0 & -28/3 & 4/3  & 0   & 0 \\
0 & 16/3  & 32/3 & 0   & 0 \\
0 & 0     & 0    & -16 & 0 \\
0 & 0     & 0    & -6  & 2 \\
\end{array}
\right)~,
\label{g0-matrix} 
\eeq 
in agreement with eqs.~(11)-(13) of ref.~\cite{Bagger}.

The renormalization group equation for the strong coupling constant in the
MSSM reads
\beq 
\be^{\rm MSSM}(\alpha_s) = \frac{d \alpha_s}{d \ln \mu^2} = -
\frac{\alpha_s^2}{4\pi} \, \be_0^{\rm MSSM} + \mc{O}(\alpha_s^3)~, 
\label{RGE-g3-SUSY} 
\eeq 
with $\be_0^{\rm MSSM} = 3 N_c - N_f$. 

The scale dependence of the squark and gluino masses, $M_s$ and
$M_{\tilde{g}}$, is described instead by the equations
\beq
\ga_{M_i}(\alpha_s) = \frac{1}{M_i^2} \frac{d M_i^2}{d \ln \mu^2} = - \as
\, \ga^{(0)}_{M_i} + O(\alpha_s^2) \qquad , \qquad i={s,\tilde{g}}~,
\label{RGE-Mi} 
\eeq 
where $\ga^{(0)}_{M_s} = 4 C_F M_{\tilde{g}}^2/M_s^2$ and $\ga^{(0)}_
{M_{\tilde g}} = 2 \be_0^{\rm MSSM} = 2(3N_c-N_f)$. 

Finally, the running of the dimensionful mass insertions $\Delta_{X}$ is
expressed by
\beqn 
&&\frac{d \Delta_{LL(RR)}}{d \ln \mu^2} = 0 + O(\alpha_s^2)~,\nn \\
&&\frac{d \Delta_{LR(RL)}}{d \ln \mu^2} = - \as \,
\ga^{(0)}_{\Delta}\Delta_{LR(RL)} + O(\alpha_s^2)
\label{D-running} 
\eeqn
with $\ga^{(0)}_{\Delta}= 2 C_F$.

By using the results given in eqs.~(\ref{RGE-g3-SUSY})-(\ref{D-running}), we
have then verified that our expressions for the Wilson coefficients exhibit
at the NLO the correct renormalization scale dependence predicted by
eq.~(\ref{chain-mu-dep}).

\subsection{Discussion of the results}
We conclude this section by presenting and discussing the final results
obtained for the Wilson coefficients at the NLO. The complete expressions of
these coefficients, in the $\MSbar$-DRED renormalization scheme, are
collected in appendix~\ref{app:coe}.

\begin{figure}[t]
\begin{center}
\includegraphics[scale=0.4,angle=270]{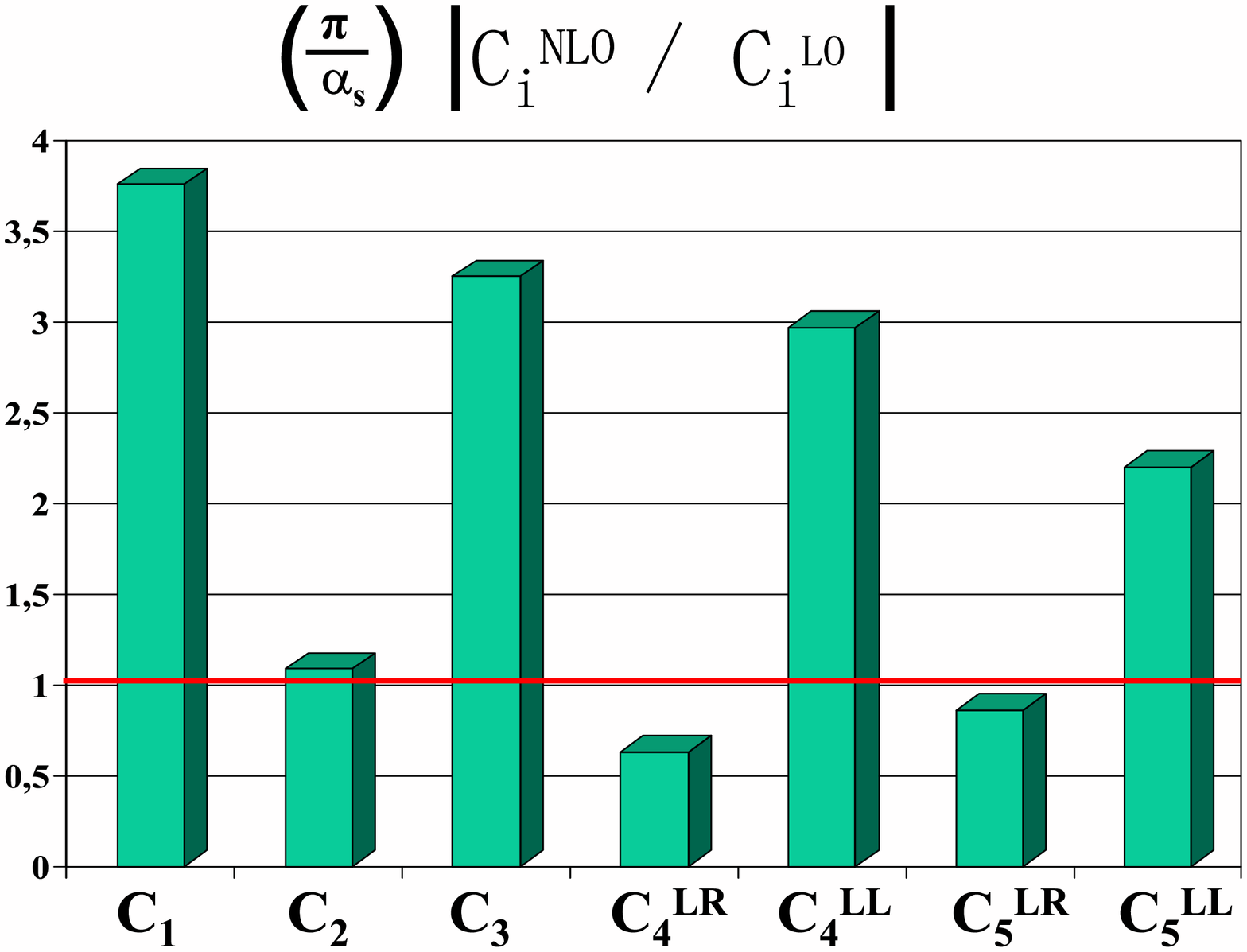}
\caption{\small\sl Comparison between the LO and NLO contributions to the 
$\MSbar$-DRED Wilson coefficients at the scale $\mu=M_s$ and at the
reference value $M_{\tilde g}=M_s$. For the coefficients $C_4$ and $C_5$ the
contributions proportional to $\delta_{LL} \delta_{RR}$ and $\delta_{LR}
\delta_{RL}$ are shown separately.}
\label{fig:cnlo}
\end{center}
\end{figure}

In order to illustrate the typical size of the computed NLO corrections, we
show in fig.~\ref{fig:cnlo} the values of the NLO contributions to the
Wilson coefficients normalized to their expected size, namely the
corresponding LO coefficients multiplied by $\alpha_s(M_s)/\pi$. For the
purpose of illustration, in this comparison we set the scale $\mu=M_s$
and put $M_{\tilde g}=M_s$.
As can be seen from the plot, in several cases the NLO coefficients turn out
to be larger than what naively expected. Of course, this conclusion applies
to the $\MSbar$-DRED coefficients and could change in a different
renormalization scheme.

\begin{figure}[p]
\begin{picture}(0,0)%
\includegraphics{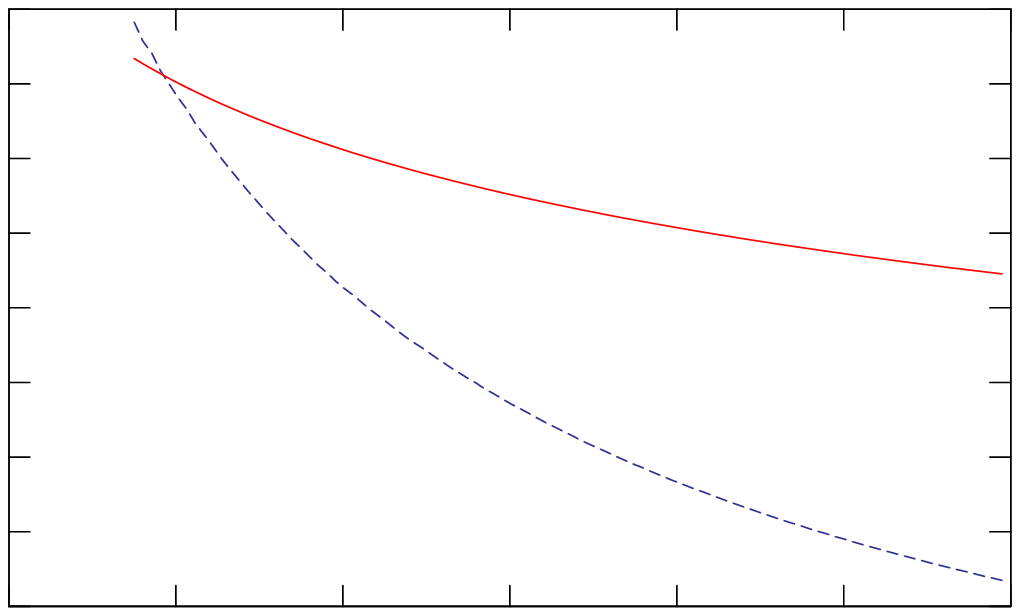}%
\end{picture}%
\begingroup
\setlength{\unitlength}{0.0200bp}%
\begin{picture}(18000,10800)(0,0)%
\put(2475,1650){\makebox(0,0)[r]{\strut{} 0.4}}%
\put(2475,2725){\makebox(0,0)[r]{\strut{} 0.42}}%
\put(2475,3800){\makebox(0,0)[r]{\strut{} 0.44}}%
\put(2475,4875){\makebox(0,0)[r]{\strut{} 0.46}}%
\put(2475,5950){\makebox(0,0)[r]{\strut{} 0.48}}%
\put(2475,7025){\makebox(0,0)[r]{\strut{} 0.5}}%
\put(2475,8100){\makebox(0,0)[r]{\strut{} 0.52}}%
\put(2475,9175){\makebox(0,0)[r]{\strut{} 0.54}}%
\put(2475,10250){\makebox(0,0)[r]{\strut{} 0.56}}%
\put(2750,1100){\makebox(0,0){\strut{} 100}}%
\put(5154,1100){\makebox(0,0){\strut{} 200}}%
\put(7558,1100){\makebox(0,0){\strut{} 300}}%
\put(9963,1100){\makebox(0,0){\strut{} 400}}%
\put(12367,1100){\makebox(0,0){\strut{} 500}}%
\put(14771,1100){\makebox(0,0){\strut{} 600}}%
\put(17175,1100){\makebox(0,0){\strut{} 700}}%
\put(550,5950){\rotatebox{90}{\makebox(0,0){\strut{}$\Delta m_d$
[ps$^{-1}$]}}}%
\put(9962,275){\makebox(0,0){\strut{}$\mu$ [GeV]}}%
\put(9482,3800){\makebox(0,0)[l]{\strut{}LO}}%
\put(13569,8100){\makebox(0,0)[l]{\strut{}NLO}}%
\put(5154,3800){\makebox(0,0)[l]{\strut{}$\left(\delta^{d}_{13}\right)_{LL,
RR}$}}%
\end{picture}%
\endgroup
\\
\begin{picture}(0,0)%
\includegraphics{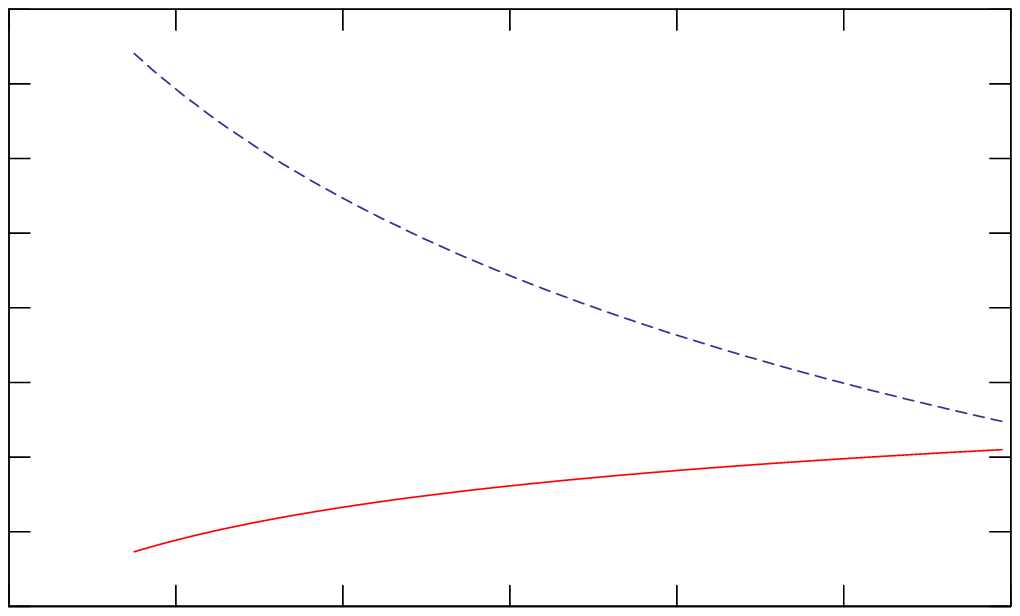}%
\end{picture}%
\begingroup
\setlength{\unitlength}{0.0200bp}%
\begin{picture}(18000,10800)(0,0)%
\put(2475,1650){\makebox(0,0)[r]{\strut{} 0.48}}%
\put(2475,2725){\makebox(0,0)[r]{\strut{} 0.5}}%
\put(2475,3800){\makebox(0,0)[r]{\strut{} 0.52}}%
\put(2475,4875){\makebox(0,0)[r]{\strut{} 0.54}}%
\put(2475,5950){\makebox(0,0)[r]{\strut{} 0.56}}%
\put(2475,7025){\makebox(0,0)[r]{\strut{} 0.58}}%
\put(2475,8100){\makebox(0,0)[r]{\strut{} 0.6}}%
\put(2475,9175){\makebox(0,0)[r]{\strut{} 0.62}}%
\put(2475,10250){\makebox(0,0)[r]{\strut{} 0.64}}%
\put(2750,1100){\makebox(0,0){\strut{} 100}}%
\put(5154,1100){\makebox(0,0){\strut{} 200}}%
\put(7558,1100){\makebox(0,0){\strut{} 300}}%
\put(9963,1100){\makebox(0,0){\strut{} 400}}%
\put(12367,1100){\makebox(0,0){\strut{} 500}}%
\put(14771,1100){\makebox(0,0){\strut{} 600}}%
\put(17175,1100){\makebox(0,0){\strut{} 700}}%
\put(550,5950){\rotatebox{90}{\makebox(0,0){\strut{}$\Delta m_d$
[ps$^{-1}$]}}}%
\put(9962,275){\makebox(0,0){\strut{}$\mu$ [GeV]}}%
\put(13569,9175){\makebox(0,0)[l]{\strut{}$\left(\delta^{d}_{13}\right)_{LR,
RL}$}}%
\put(13569,5950){\makebox(0,0)[l]{\strut{}LO}}%
\put(13569,2725){\makebox(0,0)[l]{\strut{}NLO}}%
\end{picture}%
\endgroup
\caption{\small\sl LO and NLO predictions for the $B_d$-mesons mass
difference $\Delta m_d$ obtained by adding the SUSY contributions
proportional to $\delta_{LL},\delta_{RR}$ (top) and
$\delta_{LR},\delta_{RL}$ (bottom) to the SM prediction. See text for the
reference values of the input parameters. The results are plotted as
functions of the matching scale $\mu$.}
\label{fig:scale}
\end{figure}

The Wilson coefficients depend on the matching scale $\mu$ which can be
chosen around a typical SUSY scale, \emph{e.g.}~the average squark mass
$M_s$. An important achievement of the NLO calculation is a significant
reduction of this dependence with respect to the LO approximation. This is
illustrated in fig.~\ref{fig:scale} where we show the LO and NLO predictions
for the $B_d$-mesons mass difference $\Delta m_d$ as a function of the
high-energy scale $\mu$ chosen for the matching. These predictions are
obtained by adding the SUSY contributions to the reference SM value, $\Delta
m_d^{\rm SM}=0.496$ ps$^{-1}$. The hadronic matrix elements are evaluated by
using the lattice QCD results of ref.~\cite{Bparams-lattice-DB} for the
B-parameters and $f_{B_d}=189$ MeV. We set $M_s=M_{\tilde g}=350$ GeV and
consider two cases for mass insertion coefficients, $\delta_{LL} =
\delta_{RR} = 0.12 \exp[-i \,23^{\rm o}]$ (upper plot in
fig.~\ref{fig:scale}) and $\delta_{LR} = \delta_{RL} =0.026 \exp[-i\,
23^{\rm o}]$ (lower plot in fig.~\ref{fig:scale}), chosen to give a SUSY
contribution compatible with the present measurement taking into account the
SM uncertainty. Clearly, the reduction of the scale dependence found at
the NLO quantitatively depends on the specific values chosen for the mass
insertion parameters.

From fig.~\ref{fig:scale} we see that the SUSY prediction for $\Delta m_d$
varies, at the LO, by approximately $\pm 16\%$ ($\pm 8\%$) in the LL/RR
(LR/RL) case, when the scale $\mu$ is varied in the typical range between
$M_s/2$ and $2 M_s$. With the NLO calculation, the dependence on the
matching scale is reduced by a factor two or more, \ie at the level of
$\pm 5\%$ ($\pm 2\%$) percent.

We conclude this section by observing that phenomenological applications
require the knowledge of the hadronic matrix elements. These are usually
computed on the lattice where, in order to perform a fully non-perturbative
renormalization, the RI-MOM renormalization scheme~\cite{RI-MOM} is needed.
This is the scheme adopted for instance in
refs.~\cite{Bparams-lattice-DS1,Bparams-lattice-DS2} and
\cite{Bparams-lattice-DB}, where lattice results for the complete basis of
four-fermion operator matrix elements relevant for $K-\ov{K}$ and
$B_{d,s}-\ov{B}_{d,s}$ systems have been presented. In these cases, the
results for the Wilson coefficients given in appendix~\ref{app:coe} in the
$\MSbar$-DRED scheme must be converted to the RI-MOM scheme. This can be
easily done using the relation analogous to eq.~(\ref{NDRtoDRED}), namely
\beq
C_i^{\rm RI} = \sum_j \left(1+\as \Delta r^{\rm DRED/RI}  \right)^{\mathrm
T}_{ij} C_j^{\rm \msbar-DRED} ~.
\label{DREDtoRI}
\eeq
The matrix $(\Delta r^{\rm DRED/RI})^{\mathrm T}$ which performs the
matching between the two schemes can be obtained by transposing the matrix
in eq.~(\ref{Dr-dredri}).

\section{Conclusions\label{sec:conclusions}}
\setcounter{equation}{0}
In this work we have computed the NLO strong interaction corrections to the
Wilson coefficients relevant for $\Delta F=2$ transitions in the MSSM with
the mass insertion approximation. The complete expressions for the
coefficients are given in appendix~\ref{app:coe} in the $\MSbar$-DRED
scheme. We also give in eqs.~(\ref{NDRtoDRED}) and (\ref{DREDtoRI}) the
formulae required to translate the Wilson coefficients at the NLO from the
DRED to the NDR and the RI-MOM renormalization schemes, which might be
useful for phenomenological applications.

Theoretically, the NLO calculation of the Wilson coefficients is required to
cancel the corresponding renormalization scale and scheme dependence of the
renormalized operators. Once combined with the NLO anomalous dimension of
the four-fermion operators given in ref.~\cite{NLOADM}, our results allow to
perform a complete NLO analysis of $\Delta F=2$ transitions in the MSSM. The
phenomenological analysis will be presented in a forthcoming publication.
In this study we have shown that, by considering as a reference example the
theoretical prediction of the $B_d$-meson mass difference $\Delta m_d$, the
uncertainty due to the choice of the the high-energy matching scale is
largely reduced going from the LO to the NLO, typically from about 10-15\%
to few percent.

\section*{Acknowledgments}
We warmly thank Giuseppe Degrassi and Schedar Marchetti for valuable
discussions. D.G. acknowledges the support of Fondazione Angelo Della
Riccia, Firenze, Italy.  This work has been supported in part by the
EU network ``The quest for unification'' under the contract
MRTN-CT-2004-503369.

\appendix
\renewcommand{\theequation}{\thesection.\arabic{equation}}

\section{Results for the Wilson coefficients\label{app:coe}}
\setcounter{equation}{0}

In this appendix we collect the complete NLO expressions of the Wilson
coefficients entering the effective Hamiltonian which describes $\Delta F=2$
transitions mediated by strong interactions in the MSSM.

We consider the complete basis of four fermion operators given in
eq.~(\ref{basis}). The coefficients for the operators $\tilde{Q}_{1,2,3}$
are obtained from those of the operators $Q_{1,2,3}$ by simply exchanging $L
\leftrightarrow R$ in the mass insertion parameters. For this reason, we
will not present in the following their explicit expressions.

The Wilson coefficients are written as
\beq
C_i(\mu)=C_i^{(0)}(\mu)+C_i^{(1)}(\mu)\,,
\eeq
where $\mu$ is the scale used in the matching procedure. To simplify the
notation, in the following, we will not write explicitly the $\mu$
dependence of masses and couplings.

The LO calculation of the Wilson coefficients has been performed in
ref.~\cite{GGMS} and we agree with their results. These coefficients read
\begin{footnotesize}
\beqn
&& C_1^{(0)}(\mu) = 
\frac{\alpha_{s}^{2}}{(1-x)^5\,M_s^2}
\Big[\frac{11}{108}+\frac{133\, x}{108}-\frac{13\,
{x^2}}{12}-\frac{29\, {x^3}}{108}+\frac{{x^4}}{54}+\big(\frac{13\,
x}{18}+\frac{17\, {x^2}}{18}\big)\, \log x\Big]\, \delta_{{LL}}^{2} \nn \\
&& C_2^{(0)}(\mu) = 
\frac{\alpha_{s}^{2}}{(1-x)^5\,M_s^2}
\Big[\frac{289\, x}{108}-\frac{17\, {x^2}}{12}-\frac{17\,
{x^3}}{12}+\frac{17\, {x^4}}{108}+\big(\frac{17\, x}{18}+\frac{17\,
{x^2}}{6}\big)\, \log x\Big]\, \delta_{{RL}}^{2} \nn \\
&& C_3^{(0)}(\mu) = 
\frac{\alpha_{s}^{2}}{(1-x)^5\,M_s^2}
\Big[-\frac{17\,x}{36}+\frac{{x^2}}{4}+\frac{{x^3}}{4}-\frac{{x^4}}{36}
-\big(\frac{x}{6}+\frac{{x^2}}{2}\big)\, \log x\Big]\,\delta _{{RL}}^{2}
\nn \\
&& C_4^{(0)}(\mu) =
\frac{\alpha_{s}^{2}}{{{(1-x)}^5}\,M_s^2}
\Big[\Big(-\frac{11}{54}-\frac{11\,x}{6} +\frac{11\, {x^2}}{6}+\frac{11\,
{x^3}}{54}-\frac{11\,x}{9}\big(1+x\big)\, \log x\Big)\,
{{\delta }_{{LR}}}\, {{\delta }_{{RL}}}+  \nn \\
&& \qquad \Big(-\frac{1}{9}+\frac{101\, x}{18}-\frac{5\,
{x^2}}{2}-\frac{61\, {x^3}}{18} +\frac{7\, {x^4}}{18}+\Big(\frac{5\,
x}{3}+\frac{19\, {x^2}}{3}\Big)\, \log x\Big)\, {{\delta }_{{LL}}}\,
{{\delta }_{{RR}}}\Big]  \nn \\
&& C_5^{(0)}(\mu) = \frac{\alpha_{s}^{2}}{(1-x)^5\,M_s^2}
\Big[\Big(-\frac{5}{18}-\frac{5\,x}{2} +\frac{5\, {x^2}}{2}+\frac{5\,
{x^3}}{18}-\frac{5\,x}{3}\big(1+x\big)\, \log x\Big)\,
{{\delta }_{{LR}}}\, {{\delta }_{{RL}}}+  \nn \\
&& \qquad \Big(\frac{5}{27}+\frac{107\, x}{54}-\frac{11\,
{x^2}}{6}-\frac{19\, {x^3}}{54} +\frac{{x^4}}{54}+\Big(\frac{11\,
x}{9}+\frac{13\, {x^2}}{9}\Big)\, \log x\Big)\, {{\delta }_{{LL}}}\,
{{\delta }_{{RR}}}\Big]
\label{wcLO}
\eeqn
\end{footnotesize}
where $x=M_{\tilde{g}}^2/M_s^2$ with $M_{\tilde{g}}$ the gluino mass and
$M_s$ the average squark mass. The dimensionless mass insertion parameters
are understood to be $\delta_{12}^d$, $\delta_{13}^d$, $\delta_{23}^d$ and
$\delta_{12}^u$ for the cases of $K$, $B_d$, $B_s$ and $D$ mixings
respectively.

At the NLO, the Wilson coefficients are scheme dependent quantities. Here we
present the results for the operators renormalized in the $\MSbar$-DRED
scheme, and we adopt the same scheme also for the strong coupling constant
$\alpha_s(\mu)$ and for the squark and gluino masses. Eq.~(\ref{DREDtoRI})
can be then used to convert the coefficients to the RI-MOM scheme,
frequently adopted in lattice QCD calculations of the corresponding hadronic
matrix elements. As for the strong coupling constant and the squark and
gluino masses, they can be converted to their counterparts in the
$\MSbar$-NDR scheme by using~\cite{Alta,MV} 
\beqn
&& \alpha_s^{\rm NDR} = \alpha_s^{\rm DRED} \left(1 -
\as\frac{N_c}{3}\right) \nn \\
\label{asdred}
&& M^{\rm NDR}_{\tilde{g}} = M^{\rm DRED}_{\tilde{g}} \left( 1 + \as N_c
\right) \\
&& M^{\rm NDR}_s = M^{\rm DRED}_s \left( 1 + \mc{O}(\alpha_s^2) \right) \nn
~.
\eeqn

We now present the NLO expressions for the Wilson coefficients. In the
following, the symbol $\rm{Li}_2(x)$ denotes the dilogarithm function
defined as
\beq
\rm{Li}_2(x)=-\int_0^x \frac{dt}{t} \ln(1-t)~
\eeq
and all couplings and masses are understood to be renormalized at the same
scale $\mu$. The coefficients read 
\begin{footnotesize}
\beqn
&& C_1^{(1)}(\mu) = \frac{\alpha_s^3}{\pi M_s^2}\ \frac{1}{(1-x)^7}\cdot
\nn \\ && \qquad
{\delta_{LL}}^2\,\Big[ \frac{4171}{864} + \frac{50197\,x}{2592} - 
    \frac{9911\,x^2}{144} + \frac{25039\,x^3}{432} - 
    \frac{27371\,x^4}{2592} - \frac{317\,x^5}{96} + \frac{7\,x^6}{12} - 
    \frac{4\,x^7}{81} + 
\nn \\ && \qquad
    \Big( \frac{55\,x}{9} + \frac{17005\,x^2}{1296} - 
       \frac{3607\,x^3}{144} + \frac{4319\,x^4}{1296} + 
       \frac{3875\,x^5}{1296} - \frac{5\,x^6}{9} + \frac{4\,x^7}{81} \
\Big) \,\log x + 
\nn \\ && \qquad
\Big( -\frac{247\,x}{72} - \frac{1079\,x^2}{72} + 
       \frac{3721\,x^3}{216} + \frac{497\,x^4}{216} \Big) \,\log^2 x +
\nn \\ && \qquad
\Big( -\frac{229}{108}  + 
       \frac{691\,x}{108} + \frac{67\,x^2}{36} - \frac{1559\,x^3}{108} + 
       \frac{224\,x^4}{27} \Big) \,{\rm Li}_2(1 - x) + 
\nn \\ && \qquad
    \log \Big(\frac{M_s^2}{\mu^2}\Big)\,
     \Big( -\frac{11}{54}  - \frac{577\,x}{108} + 
       \frac{1094\,x^2}{81} - \frac{2389\,x^3}{324} - 
       \frac{1255\,x^4}{324} + \frac{310\,x^5}{81} - \frac{191\,x^6}{324} + 
       \frac{4\,x^7}{81} + 
\nn \\ && \qquad
\Big( -\frac{91\,x}{36} - 
          \frac{133\,x^2}{108} + \frac{935\,x^3}{108} - \frac{529\,x^4}{108}
 \Big) \,\log x \Big)  \Big]
\eeqn

\beqn
&& C_2^{(1)}(\mu) = \frac{\alpha_s^3}{\pi M_s^2}\ \frac{1}{(1-x)^7}\cdot
\nn \\ && \qquad
{\delta_{RL}}^2\,\Big[ \frac{311827\,x}{3888} - 
    \frac{783947\,x^2}{3888} + \frac{308057\,x^3}{1944} - 
    \frac{58433\,x^4}{1944} - \frac{39589\,x^5}{3888} + 
    \frac{14093\,x^6}{3888} - \frac{34\,x^7}{81} + 
\nn \\ && \qquad
    \Big( \frac{107\,x}{9} + \frac{2285\,x^2}{108} - 
       \frac{5839\,x^3}{162} - \frac{109\,x^4}{81} + \frac{41\,x^5}{6} - 
       \frac{943\,x^6}{324} + \frac{34\,x^7}{81} \Big) \,\log x + 
\nn \\ && \qquad
    \Big( -\frac{199\,x}{216} - \frac{7535\,x^2}{216} + 
       \frac{5869\,x^3}{216} + \frac{803\,x^4}{72} \Big) \,\log^2 x + 
\nn \\ && \qquad
    \Big( -\frac{191\,x}{18} + \frac{2011\,x^2}{54} - 
       \frac{2303\,x^3}{54} + \frac{865\,x^4}{54} \Big) \,
     {\rm Li}_2(1 - x) + 
\nn \\ && \qquad
    \log \Big(\frac{M_s^2}{\mu^2}\Big)\,
     \Big( -\frac{6205\,x}{648} + \frac{10891\,x^2}{648} + 
       \frac{2077\,x^3}{324} - \frac{9035\,x^4}{324} + 
       \frac{11411\,x^5}{648} - \frac{2453\,x^6}{648} + \frac{34\,x^7}{81}+
\nn \\ && \qquad
       \Big( -\frac{311\,x}{108} - \frac{991\,x^2}{108} + 
          \frac{2575\,x^3}{108} - \frac{1273\,x^4}{108} \Big) \,\log x \
\Big)  \Big]
\eeqn

\beqn
&& C_3^{(1)}(\mu) = \frac{\alpha_s^3}{\pi M_s^2}\ \frac{1}{(1-x)^7}\cdot
\nn \\ && \qquad
{\delta_{RL}}^2\,\Big[ -\frac{43993\,x}{3888} + 
    \frac{103649\,x^2}{3888} - \frac{35339\,x^3}{1944} + 
    \frac{2147\,x^4}{1944} + \frac{9103\,x^5}{3888} - 
    \frac{2663\,x^6}{3888} + \frac{2\,x^7}{27} + 
\nn \\ && \qquad
    \Big( \frac{43\,x}{27} - \frac{3677\,x^2}{324} + 
       \frac{1537\,x^3}{162} + \frac{59\,x^4}{27} - \frac{401\,x^5}{162} + 
       \frac{205\,x^6}{324} - \frac{2\,x^7}{27} \Big) \,\log x + 
\nn \\ && \qquad
    \Big( \frac{277\,x}{216} + \frac{55\,x^2}{8} - \frac{175\,x^3}{24} - 
       \frac{283\,x^4}{216} \Big) \,\log^2 x + 
\nn \\ && \qquad
    \Big( \frac{71\,x}{54} - \frac{385\,x^2}{54} + \frac{557\,x^3}{54} - 
       \frac{9\,x^4}{2} \Big) \,{\rm Li}_2(1 - x) + 
\nn \\ && \qquad
    \log \Big(\frac{M_s^2}{\mu^2}\Big)\,
     \Big( \frac{1607\,x}{648} - \frac{3217\,x^2}{648} + 
       \frac{25\,x^3}{324} + \frac{1745\,x^4}{324} - \frac{2345\,x^5}{648}
+ 
       \frac{463\,x^6}{648} - \frac{2\,x^7}{27} + 
\nn \\ && \qquad
       \Big( \frac{85\,x}{108} + \frac{205\,x^2}{108} - 
          \frac{605\,x^3}{108} + \frac{35\,x^4}{12} \Big) \,\log x \
\Big)  \Big] 
\eeqn

\beqn
&& C_4^{(1)}(\mu) = \frac{\alpha_s^3}{\pi M_s^2}\ \frac{1}{(1-x)^7}\cdot
\Big\{
\nn \\ && \qquad
\delta_{LL}\,\delta_{RR}\,
   \Big[ - \frac{203}{36}  + \frac{39341\,x}{216} - 
     \frac{91895\,x^2}{216} + \frac{33707\,x^3}{108} - 
     \frac{733\,x^4}{18} - \frac{6691\,x^5}{216} + \frac{2069\,x^6}{216} - 
     \frac{28\,x^7}{27} + 
\nn \\ && \qquad
     \Big( \frac{31\,x}{12} + \frac{8435\,x^2}{108} - 
        \frac{7927\,x^3}{108} - \frac{2065\,x^4}{108} + 
        \frac{490\,x^5}{27} - \frac{397\,x^6}{54} + \frac{28\,x^7}{27} \
\Big) \,\log x + 
\nn \\ && \qquad
\Big( -\frac{43\,x}{8} - \frac{1873\,x^2}{24} + 
        \frac{4687\,x^3}{72} + \frac{1703\,x^4}{72} \Big) \,\log^2 x\ + 
\nn \\ && \qquad
\Big( \frac{49}{18} - \frac{721\,x}{18} + 
        \frac{599\,x^2}{6} - \frac{1627\,x^3}{18} + \frac{251\,x^4}{9} \
\Big) \,{\rm Li}_2(1 - x) + 
\nn \\ && \qquad
\log \Big(\frac{M_s^2}{\mu^2}\Big)\,\Big( \frac{1}{12} -
\frac{461\,x}{72} - 
        \frac{413\,x^2}{216} + \frac{4529\,x^3}{108} - 
        \frac{1600\,x^4}{27} + \frac{7109\,x^5}{216} - 
        \frac{1813\,x^6}{216} + \frac{28\,x^7}{27} + 
\nn \\ && \qquad
        \Big( -\frac{3\,x}{4} - \frac{607\,x^2}{36} + 
           \frac{1055\,x^3}{36} - \frac{421\,x^4}{36} \Big) \,\log x \
\Big)  \Big]  + 
\nn \\ && \qquad
\delta_{LR}\,\delta_{RL}\,
   \Big[ - \frac{15031}{1296} - \frac{6875\,x}{1296} + 
     \frac{33709\,x^2}{648} - \frac{20747\,x^3}{648} - 
     \frac{16963\,x^4}{1296} + \frac{13297\,x^5}{1296} - 
     \frac{22\,x^6}{81} + 
\nn \\ && \qquad
     \Big( \frac{163\,x}{36} - \frac{26977\,x^2}{648} + 
        \frac{25831\,x^3}{648} + \frac{359\,x^4}{216} - 
        \frac{3041\,x^5}{648} + \frac{22\,x^6}{81} \Big) \,\log x + 
\nn \\ && \qquad
     \Big( \frac{103\,x}{8} + \frac{703\,x^2}{72} - 
        \frac{4969\,x^3}{216} - \frac{91\,x^4}{72} \Big) \,\log^2 x + 
\nn \\ && \qquad
\Big( \frac{385}{108} - \frac{217\,x}{54} - 
        \frac{13\,x^2}{3} + \frac{349\,x^3}{54} - \frac{181\,x^4}{108} \
\Big) \,{\rm Li}_2(1 - x) + 
\nn \\ && \qquad
     \log \Big(\frac{M_s^2}{\mu^2}\Big)\,
      \Big( - \frac{25}{648}  + \frac{3257\,x}{648} - 
        \frac{3481\,x^2}{324} + \frac{1369\,x^3}{324} + 
        \frac{2683\,x^4}{648} - \frac{1867\,x^5}{648} + \frac{22\,x^6}{81}+
\nn \\ && \qquad
        \Big( \frac{173\,x}{108} + \frac{311\,x^2}{108} - 
           \frac{307\,x^3}{36} + \frac{437\,x^4}{108} \Big) \,\log x \
\Big)  \Big] \Big\}
\eeqn

\beqn
&& C_5^{(1)}(\mu) = \frac{\alpha_s^3}{\pi M_s^2}\ \frac{1}{(1-x)^7}\cdot
\Big\{
\nn \\ && \qquad
\delta_{LL}\,\delta_{RR}\,
   \Big[ \frac{305}{36} + \frac{6521\,x}{216} - \frac{72913\,x^2}{648} + 
     \frac{31549\,x^3}{324} - \frac{3193\,x^4}{162} - 
     \frac{2885\,x^5}{648} + \frac{451\,x^6}{648} - \frac{4\,x^7}{81} + 
\nn \\ && \qquad
     \Big( \frac{361\,x}{36} + \frac{1775\,x^2}{108} - 
        \frac{10505\,x^3}{324} + \frac{929\,x^4}{324} + 
        \frac{100\,x^5}{27} - \frac{107\,x^6}{162} + \frac{4\,x^7}{81} \
\Big) \,\log x + 
\nn \\ && \qquad
\Big( -\frac{407\,x}{72} - \frac{653\,x^2}{24} + 
        \frac{6361\,x^3}{216} + \frac{1121\,x^4}{216} \Big) \,
      \log^2 x + 
\nn \\ && \qquad
     \Big( -\frac{209}{54}  + 
\frac{641\,x}{54} - 
        \frac{7\,x^2}{18} - \frac{1045\,x^3}{54} + \frac{317\,x^4}{27} \
\Big) \,{\rm Li}_2(1 - x) + 
\nn \\ && \qquad
\log \Big(\frac{M_s^2}{\mu^2}\Big)\,
      \Big( -\frac{35}{108}  - \frac{1787\,x}{216} + 
        \frac{13765\,x^2}{648} - \frac{4057\,x^3}{324} - 
        \frac{370\,x^4}{81} + \frac{3299\,x^5}{648} - 
        \frac{451\,x^6}{648} + \frac{4\,x^7}{81} + 
\nn \\ && \qquad
        \Big( -\frac{143\,x}{36} - \frac{169\,x^2}{108} + 
           \frac{1385\,x^3}{108} - \frac{787\,x^4}{108} \Big) \,\log x \
\Big)  \Big]  + 
\nn \\ && \qquad
\delta_{LR}\,\delta_{RL}\,
   \Big[ -\frac{5425}{432}  - \frac{12125\,x}{432} + 
     \frac{26875\,x^2}{216} - \frac{24125\,x^3}{216} + 
     \frac{10715\,x^4}{432} + \frac{1495\,x^5}{432} - \frac{10\,x^6}{27} + 
\nn \\ && \qquad
     \Big( -\frac{155\,x}{12} - \frac{5095\,x^2}{216} + 
        \frac{10465\,x^3}{216} - \frac{655\,x^4}{72} - 
        \frac{695\,x^5}{216} + \frac{10\,x^6}{27} \Big) \,\log x + 
\nn \\ && \qquad
     \Big( \frac{75\,x}{8} + \frac{625\,x^2}{24} - \frac{2455\,x^3}{72} - 
        \frac{85\,x^4}{24} \Big) \,\log^2 x + 
\nn \\ && \qquad
     \Big( \frac{175}{36} - \frac{295\,x}{18} + 5\,x^2 + 
        \frac{355\,x^3}{18} - \frac{475\,x^4}{36} \Big) \,
      {\rm Li}_2(1 - x) + 
\nn \\ && \qquad
     \log \Big(\frac{M_s^2}{\mu^2}\Big)\,
      \Big( \frac{185}{216} + \frac{2855\,x}{216} - \frac{4135\,x^2}{108}
+ 
        \frac{3175\,x^3}{108} - \frac{155\,x^4}{216} - 
        \frac{1045\,x^5}{216} + \frac{10\,x^6}{27} + 
\nn \\ && \qquad
        \Big( \frac{275\,x}{36} - \frac{55\,x^2}{36} - 
           \frac{205\,x^3}{12} + \frac{395\,x^4}{36} \Big) \,\log x \
\Big)  \Big] \Big\}
\eeqn
\end{footnotesize}
\bibliography{biblio} 
\bibliographystyle{JHEP}

\end{document}